\documentclass[twocolumn,dvipsnames]{autart}    

\usepackage{graphicx}          
\usepackage[authoryear,round]{natbib}        
\usepackage{ifthen}
\usepackage{amsmath,amssymb}
\usepackage{siunitx}
\usepackage{booktabs}
\usepackage{multirow}
\usepackage{tikz}
\usepackage{subfigure}
\usetikzlibrary{shapes.misc}

\usepackage{auto18cmds}
\usepackage{xcolor}
\usepackage{tikz}
\usetikzlibrary{shapes.misc,math,calc}

\usepackage{pgfplots}
\pgfplotsset{compat=newest}
\usepgfplotslibrary{external}
\begin{document}

\begin{frontmatter}

\title{PDE-based multi-agent formation control using flatness and
  backstepping: analysis, design and robot experiments} 


\author{Gerhard Freudenthaler},    
\author{Thomas Meurer}             

\address{\{gefr,tm\}@tf.uni-kiel.de,\\Chair of Automatic Control, Faculty of Engineering, Kiel University}  


\begin{abstract}                          
  A PDE-based control concept is developed to deploy a multi-agent
  system into desired formation profiles.
  The dynamic model is based on a coupled linear,
  time-variant parabolic distributed parameter system. By means of a
  particular coupling structure parameter information can be
  distributed within the agent continuum.
  Flatness-based motion planning and feedforward control are combined with a
  backstepping-based boundary controller to stabilise the distributed
  parameter system of the tracking error. The tracking controller
  utilises the required state information from a 
  Luenberger-type state observer. By means of an exogenous
  system the relocation of formation profiles is achieved. The transfer
  of the control strategy to a finite-dimensional discrete multi-agent system is
  obtained by a suitable finite difference discretization of the continuum
  model, which in addition imposes a leader-follower communication
  topology. The results are evaluated both in simulation studies and
  in experiments for a swarm of mobile robots realizing the transition
  between different stable and unstable formation profiles.
\end{abstract}

\end{frontmatter}


\def\pdes{PDEs}
\def\twodof{2DOF}
\def\fdfd{feedforward}
\def\PDE{PDE}
\def\mas{multi-agent system }
\def\MAS{MAS}

\tikzset{cross/.style={line width=1pt,cross out, draw=black, minimum size=2*(#1-\pgflinewidth), inner sep=0pt, outer sep=0pt},
  cross/.default={1pt}}

%
\section{Introduction}\label{sec:intro}
%
In general multi-agent systems consist of interconnected dynamic
subsystems which share information. This elementary concept opens up a
wide field of applications such as consensus and synchronisation
problems, decision making, crowd dynamics, formation control, cooperative
multi-vehicle control, or complex oscillators networks
\citep{olfati-saber_consensus_2007,murray_recent_2007,mesbahi_graph_2010,easley_kleinberg_2010,dorfler_synchronization_2014,bullo:18}. 

Different approaches have been proposed to model and to control the
dynamic behaviour of multi-agent systems. Behaviour-based approaches are discussed in
\cite{reynolds_flocks_1987, balch_behavior-based_1998} while
\cite{leonard_virtual_2001} make use of artificial potential and
virtual leaders. Graph theory is a widespread concept to model the
behaviour of multi-agent system, complex networks, or swarms by
studying ODE representations
\citep{olfati-saber_consensus_2007}. However, continuum models in terms
of partial differential equations (\pdes) have increasingly been used
to describe the dynamics of many interacting participants, see, e.g.,
\cite{ frihauf_leader-enabled_2011, meurer_finite-time_2011,
  meurer_control_2013, qi_multi-agent_2015,
  pilloni_consensus-based_2016, freudenthaler_pde-based_2016,
  freudenthaler_backsteppingbased_2017}. The motivation comes from the
fact, that certain semi-discretised PDEs match the pattern of
important graph-based dynamic models, e.g., the Graph-Laplacian consensus
protocol. This can be exploited
to develop PDE-based control and estimation algorithms. Following this
process consisting of (i) first imposing a desired PDE dynamics for the
multi-agent continuum, (ii) performing PDE based control design and then
(iii) realizing the transfer to discrete multi-agent systems by proper PDE
discretization, characterises an inverse design approach that is in
principle independent of the number of agents and their communication
topology \citep{meurer_finite-time_2011,meurer_control_2013}. 

\begin{figure}[tb]
  \begin{center}
    \includegraphics{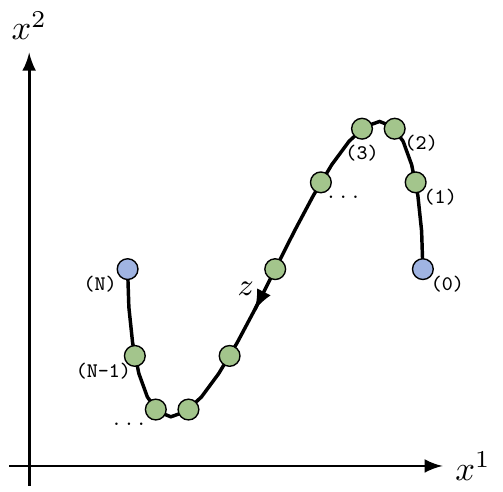}
    \label{fig:sec1:MAS_Example}
    \caption{Formation profile of $ 11 $ agents in the $ \of{x^1,x^2} $-plane;
      $ \color{NavyBlue!40} \bullet $ refer to the leader agents,
      $ \color{OliveGreen!40} \bullet $ denote the
      followers.}
  \end{center}
\end{figure}
The basic idea is illustrated schematically in Fig. 1. Herein, $ 11 $
agents in a so-called leader-follower configuration are arranged in
the plane. The two types of agents refer to active and 
collaborative roles, however leaders may have to fulfil more
sophisticated tasks than followers. By moving from the discrete set of
agents denoted by coloured dots to an agent continuum the formation is
visualized by the line with $ z $ being interpreted as a virtual
communication path. The formation is thereby obtained by the
superposition of solutions of \pdes~in the individual directions $x^1$
and $x^2$. 

This contribution addresses the design of a two-degrees-of-freedom
(\twodof) boundary control concept. The approach combines motion planning and
feedforward control with stabilising tracking control for a multi-agent
continuum model in terms of coupled linear, time-variant
diffusion-reaction equations. It is shown that this setting allows us
to recover a wide range of common multi-agent dynamics and enables us to
realise various formation shapes. This extends the previous work
\citep{freudenthaler_pde-based_2016} of the authors in several
directions: (i) a state observer for the continuum model is included
into the control loop; (ii) the stability of the closed-loop control
is rigorously assessed using Lyapunov's stability theory; (iii) the
decentralised distribution and synchronisation of in particular
parameter values through the multi-agent network is addressed; (iv) a
first experimental verification of the theoretical results is provided
using a small swarm of mobile robots. 

Motion planning and feedforward control design are based on the
flatness property of the continuum model, which is exploited by taking
into account results from
\cite{meurer_trajectory_2009,freudenthaler_pde-based_2016} to use the 
formal integration of the coupled PDEs. The continuum model is 
composed of two PDEs with the state of the first PDE referring to the
spatial location of an agent element. The second PDE couples into the
first PDE and governs the spatial-temporal distribution of its 
reaction parameter. By controlling this reaction parameter evolution
desired parameter adaptations can be conducted, which results in a rich class
of possible formation profiles. Formations herein correspond to steady
state solutions of the continuum model. 
To address the deployment into unstable formations the flatness-based
feedforward control is extended by an error state feedback to obtain a tracking
controller involving a Luenberger-type state observer. The design of
both the controller and the observer makes use of the backstepping
technique, which has been extensively studied for different types of
\pdes, see, e.g., \cite{krstic_smyshlyaev_siam:08}. For
diffusion-reaction equations the linear, time-invariant case
is addressed, e.g., in \cite{smyshlyaev_closed-form_2004,
  smyshlyaev_backstepping_2005,baccoli_boundary_2015} with extensions
to the time-varying case provided, e.g., in \cite{meurer_tracking_2009,
  jadachowski_efficient_2012,meurer_control_2013}.
In addition to simulation studies this contribution presents first
experimental results for the considered PDE-based formation control
concept by using a small swarm of mobile robots. It is shown that the
combined flatness- and backstepping-based tracking controller enables
us to experimentally achieve transitions even into unstable formation
profiles with a spatial relocation of the swarm. 

The article is organized as follows: Section~\ref{sec:problem}
introduces the model of the agent dynamics involving the
spatial-temporal parameter evolution and defines steady state
formation profiles. The \twodof~control concept is discussed in the
two subsequent sections including the flatness-based feedforward control
approach in Section~\ref{sec:flatnessFF} and the observer-based
stabilisation of the tracking error dynamics in
Section~\ref{sec:trckngCtrl}. The formal transfer to the discrete
setup imposing the communication topology and simulations studies are
provided in Section~\ref{sec:simStuds}. The implementation at a
test-rig and experimental results are presented in Section
\ref{sec:expResults}. Some final remarks in Section~\ref{sec:concl}
conclude the paper.


%
\section{Problem formulation}\label{sec:problem}
%
In the following a continuum formulation using PDEs is introduced to
model the agent dynamics. For this the connection between the
  continuum model and the related ODE formulation of a multi-agent
  system under next-neighbor communication is addressed, which is also
  utilized in Sections \ref{subsec:communication} and
  \ref{sec:expResults} for the implementation in the simulation and
  the experimental environment.
%
\subsection{Multi-agent system model}
\label{subsec:sec2:mas-model}
%
Taking into account the undirected line graph $G(V,E)$ of
Fig. \ref{fig:sec1:MAS_Example} with node set $V$ and edge set
$E$. Nodes $j,\,k\in V$ can share information if $(i,j)\in E$, i.e.,
$i\sim j$. Let
$(\sx_{j}^{1}(t),\sx_{j}^{2}(t))$ denote the position of agent $j$ at
time $t$ in the $(x^1,x^2)$-plane. Let $N+1$ denote the number
of nodes in $V$. Nodes $j\in V$ are numbered consecutively starting from $0$ to
$N$ with $j\in\{0,N\}$ denoting leaders and $j\in\{1,N-1\}$
representing followers. The multi-agent system is considered under the
(time-varying) next-neighbor protocol 
\begin{subequations}
  \label{eq:sec2:discrete:all}
  \begin{align}
    \dot{\sx}_{j}^{i}(t) &= \mathfrak{a}^{i}
                           \sum_{k\sim j} \big(\sx_{k}^{i}(t) -
                           \sx_{j}^{i}(t)\big) + c_j^{i}(t)
                           \sx_{j}^{i}(t)  \label{eq:sec2:x:discrete:ode}\\
    \dot{c}_{j}^{i}(t) &= \mathfrak{b}^{i}
                         \sum_{k\sim j} \big(c_{k}^{i}(t) -
                         c_{j}^{i}(t)\big) + d_j^{i}(t)
                         c_{j}^{i}(t)\label{eq:sec2:c:discrete:ode}
  \end{align}
  for all follower agents $j=1,2,\ldots,N-1$.
  The parameters satisfy $\mathfrak{a}^{i}>0$,
  $\mathfrak{b}^{i}>0$, $d_j^{i}\t\in\R$ with $\mathfrak{a}^{i}$,
  $\mathfrak{b}^{i}$ in general showing some proportionality to
  $1/(N+1)$ by means of the adjacency matrix \citep{mesbahi_graph_2010}
  or particular influence functions in opinion dynamics
  \citep{motsch_tadmor:14}. While \eqref{eq:sec2:x:discrete:ode}
  describes the motion of the agents the variable $c_{j}^{i}(t)$, as
  it couples into \eqref{eq:sec2:x:discrete:ode}, enables us to
  distribute parameter information, which directly influences the
  agent dynamics. It is shown subsequently that this broadens the
  applicability of the setup in particular for motion planning
  and formation control. If $d_j^{i}=0$, then this information 
    processing requires only relative data, i.e., $c_{k}^{i}(t) -
    c_{j}^{i}(t)$ for $k\sim j$. The protocol includes the graph-Laplacian control to
  achieve consensus \citep{olfati-saber_consensus_2007}.
  
  To control agent motion and parameter information external control
  signals are imposed at the leader agents $j\in\{0,N\}$ in terms of 
  \begin{align}
    & \dot{\sx}_{0}^{i}(t) = \uu[_0^i],&& \dot{\sx}_{N}^{i}(t) =
                                          \uu[_N^i]\label{eq:sec2:x:discrete:leader}\\
    & \dot{c}_{0}^{i}(t) = \vv[_0^{i}],&& \dot{c}_{N}^{i}(t) =
                                           \vv[_N^i].\label{eq:sec2:c:discrete:leader}
  \end{align}
  At the time $t=\tsymb_0$ the agents are at the initial state
  \begin{align}
    &\sx_{j}^{i}(\tsymb_0) = \sx_{j,0}^{i},&& c_{j}^{i}(\tsymb_0) = c_{j,0}^{i}.
  \end{align}
\end{subequations}
%
%
\subsection{From discrete to diffusion-like continuum model}
%
When considering a large-scale multi-agent system it is reasonable to
map the discrete agent set $j\in V$ into an agent continuum defined on the
continuous coordinate $z\in[0,\ell]$ representing the agent index in
the continuous communication topology. In view of this, the states 
$\sx_{j}^{i}(t)$ and $c_{j}^{i}(t)$ approach $\x[^{i}]$ and $c^{i}\zt$
and the next-neighbor configuration \eqref{eq:sec2:discrete:all}
translates into the coupled diffusion-reaction system (DRS) 
\begin{subequations}
  \label{eq:sec2:problem}
  \begin{align}
    \sdcr[{&}]{\x[^{i}]}{a^i}{0}{c^i\zt}\label{eq:sec2:problemPDE:x}\\
    \sdcr[{&}]{c^{i}\zt}{b^i}{0}{d^i\zt}\label{eq:sec2:problemPDE:c}
  \end{align}
  defined on the domain
  $\mathcal{D}(\ell):=\{\zt\in(0,\ell)\times\left(\tsymb_0,\infty\right)\}$
  with the boundary controls
  \begin{align}
    \sbcn{\sx^i}&= \uu[_0^i], && \sbcl{\sx^i}=\uu[_\ell^i]\label{eq:sec2:x:cont:leader}\\
    \sbcn{c^i}&= \vv[_0^i], && \sbcl{c^i}=\vv[_\ell^i]\label{eq:sec2:c:cont:leader}
  \end{align}
  and the initial conditions 
  \begin{align}\label{eq:sec2:problemIC}
    &\ic{\sx^i} = \xn[_0^i]
    &&
    \ic{c^i} = c_0^{i}(z).  
  \end{align}
\end{subequations}
\begin{rem}
  Since the problem formulation \eqref{eq:sec2:problem} is independent
  for each tuple $(\x[^{i}],c^{i}(z,t))$ the superscript $i$ referring
  to the coordinate axis is subsequently omitted.
\end{rem}
The following proposition addresses a remark by Enrique Zuazua
concerning collective dynamics using mean field and diffusion-like PDE
approaches \citep{zuazua_scindis:18}.
\begin{prop}[Discrete vs. continuum dynamics]\label{prop:sec2:discrete_vs_cont}
  Consider a next-neighbor configuration on a line graph with $N+1$
  nodes. Denote by $\Delta z=\ell/N$ the step size and introduce the
  location (agent index) $z_j=j\Delta z$ with $j=0,1,\ldots,N<\infty$.
  The discrete formulation \eqref{eq:sec2:discrete:all} and the
  continuum formulation \eqref{eq:sec2:problem} can be formally
  exchanged at any $z_j$ with error $O^4(\Delta z)$ up to the scaling
  $t\mapsto r(N,\ell) t$, $c\mapsto c/r(N,\ell)$, $d\mapsto d/r(N,\ell)$ with $r(N,\ell)=(\ell^2
  \mathfrak{a})/(N^2 a)$.
\end{prop}
The claim is supposed to provide a principal connection and makes
use of classical smooth solutions of \eqref{eq:sec2:problem} to
allow for a Taylor series expansion.
\begin{pf}
  Assuming regularity of solutions set $c_j\t=c(z_j,t)$, $\sx_{j}(t)=
  x(z_j,t)$ and consider the Taylor series expansion 
  \begin{align*}
    \sx_{j\pm 1}(t) &= x((j\pm 1)\Delta z,t)\\
    &= x(j\Delta z,t)
      \pm \partial_z x(j\Delta z,t) \Delta z\\
    &\phantom{=}\,+ \partial^{2}_z
    x(j\Delta z,t)\frac{(\Delta z)^2}{2!} \pm O^3(\Delta z).
  \end{align*}
  Replacing $x$ by $c$ yields the respective expansion for $\sc_{j\pm
    1}(t)$. Substitution into \eqref{eq:sec2:discrete:all} for a
  two-neighbor configuration yields
  \begin{subequations}
    \label{eq:sec2:ode_cont_vs_disc}
    \begin{align}
      \partial_t x(z_j,t) &= \mathfrak{a}(\Delta z)^2 \partial^{2}_z
                            x(z_j,t) + c(z_j,t) x(z_j,t) \\
      \partial_t c(z_j,t) &= \mathfrak{b}(\Delta z)^2 \partial^{2}_z
                            c(z_j,t) + d(z_j,t) c(z_j,t) 
    \end{align}
  \end{subequations}
  with error $O^4(\Delta z)$. With $\Delta z = \ell/N$ the proposed
  scaling in terms of $r(N,\ell)$ transfers
  \eqref{eq:sec2:ode_cont_vs_disc} to 
  \begin{subequations}
    \label{eq:sec2:ode_cont_vs_disc:mapped}
    \begin{align}
      \partial_t x(z_j,t) &= a \partial^{2}_z
                            x(z_j,t) + c(z_j,t) x(z_j,t) \\
      \partial_t c(z_j,t) &= b \partial^{2}_z
                            c(z_j,t) + d(z_j,t) c(z_j,t) 
    \end{align}
  \end{subequations}
  with $b= {\mathfrak{b}a}/{\mathfrak{a}}$.
  The computation above addresses the transfer from the discrete to
  the continuum model, the reverse can be obtained, e.g., by a finite
  difference discretization of \eqref{eq:sec2:problem} or similarly
  the substitution of the Taylor series expansion. 
  Comparing \eqref{eq:sec2:ode_cont_vs_disc:mapped} with
  \eqref{eq:sec2:problem} illustrates the formal relationship. Note
  that alternatively $\ell$ can be adjusted for unscaled $t$.
\end{pf}
To further interpret the result let $c_j\t=c(z_j,t)=0$. In this case
the discrete formulation is the graph-Laplace protocol
\citep{olfati-saber_flocking_2006}. Hence, as the number of nodes $N+1$
increases (thus $1/(N+1)$ decreases) all non-zero eigenvalues of the
system matrix tend to zero. For $N\to\infty$ the system rests in the
initial state. This is contrary to the dynamical behavior of the
structurally corresponding heat equation \eqref{eq:sec2:problem}
obtained for $c\zt=0$. Taking into account the time scaling $t\mapsto
(\ell^2\mathfrak{a})/(N^2 a) t$ for $N\to\infty$ reproduces the
discrete case. 
%
\subsection{Formation control problem}\label{sec:sec2:frmtnProfiles}
%
The deployment of the multi-agent system into desired
formation profiles and the finite time transition between different
formation profiles is addressed by developing a combined
feedforward-feedback control 
strategy for the leader agents \eqref{eq:sec2:x:discrete:leader},
\eqref{eq:sec2:c:discrete:leader} in the discrete setting or
\eqref{eq:sec2:x:cont:leader}, \eqref{eq:sec2:c:cont:leader} in the
continuum setting, respectively. 
\begin{defn}[Steady state]
  \label{def:sec2:steady_state}
  Let $d\zt$ be continuous in $z$, smooth in $t$ but locally
  non-analytic with $d\zt=\davr$ and $\spdt[n]d\zt = 0$ for all
  $n\geq1$ at some fixed $t$. The tuple
  \begin{align}
    \label{eq:sec2:stdyStSol:general}
    (\xavr,\cavr) = (\sxavr(\zsymb;\sxavr_{0},\sxavr_{\ell}),
    \scavr(\zsymb;\scavr_{0},\scavr_{\ell},\bar{d}\z)).
  \end{align}
  with $\xavr,\,\cavr\in C^2([0,\ell])$ is a steady state of
  \eqref{eq:sec2:problem} if
  \begin{subequations}
    \label{eq:sec2:problemStdyst}
    \begin{align}
      \label{eq:sec2:problemStdyst:ode}
      a\spddz{\xavr} + \cavr\xavr = 0,~b\spddz{\cavr} + \davr\cavr = 0
    \end{align}
    for $\zsymb\in(0,\ell)$ and 
    \begin{align}
      \label{eq:sec2:problemStdyst:bc}
      \sxavr(0) = \sxavr_{0},\quad
      \sxavr(\ell)=\sxavr_{\ell},\quad\scavr(0) = \scavr_{0},\quad
      \scavr(\ell)=\scavr_{\ell}
    \end{align}
    hold true for some $\sxavr_{0},\,\sxavr_{\ell},\,
    \scavr_{0},\,\scavr_{\ell}\in\R$.  
  \end{subequations}
\end{defn}
The constant boundary values $\sxavr_\mathrm{0}$, $\sxavr_\mathrm{\ell}$,
$\scavr_\mathrm{0}$, and $\scavr_\mathrm{\ell}$ can be freely assigned
since under steady state conditions \eqref{eq:sec2:x:cont:leader},
\eqref{eq:sec2:c:cont:leader} reduce to
$\uu[_0]=\uu[_\ell]=\vv[_0]=\vv[_\ell]=0$.
\begin{rem}
  For the sake of simplicity no distinction between different
  boundary values $\sxavr_\mathrm{0}$, $\sxavr_\mathrm{\ell}$,
  $\scavr_\mathrm{0}$, and $\scavr_\mathrm{\ell}$ is made. This is, of course,
  implicitly included. 
\end{rem}
\begin{defn}[Set of steady states]
  \label{def:sec2:set_steady_state}
  Let $d\zt$ be continuous in $z$, smooth in $t$ but locally
  non-analytic at discrete time instances $t_j$, $1\leq j<\infty$ with 
  $0\leq t_0$, $t_{j-1}<t_j$ so that
  $d(z,t_j)=\bar{d}_{t_j}\z$ and $\spdt[n]d\zt\vert_{t=t_j} = 0$ for all
  $n\geq1$. The set of steady states endowed with the $C^2$ topology
  is denoted by $\mathcal{S} = \bigcup_{j}\{(\xavr[_{t_j}],\cavr[_{t_j}])\}$
  with the tuple $(\xavr[_{t_j}],\cavr[_{t_j}])$ solving
  \eqref{eq:sec2:problemStdyst} for $\davr=\bar{d}_{t_j}\z$. 
\end{defn}
With these preparations the considered type of formation profiles and
spatial-temporal formation transitions can be properly introduced:
\begin{enumerate}
\item Formation profiles denoted by the tuple $(\xavrd,\cavrd)$ are steady
  states according to Definition \ref{def:sec2:steady_state} so that 
  \begin{align}
    \label{eq:sec2:stdyStSol}
    \xavrd\!=\!\sxavr(\zsymb;\sxavrd[_{0}],\sxavrd[_{\ell}]),~
    \cavrd\!=\!\scavr(\zsymb;\scavrd[_{0}],\scavrd[_{\ell}],\bar{d}^\ast\z).
  \end{align}
\item Denote by $(\xavrd[_{t_{j-1}}],\cavrd[_{t_{j-1}}])$
  and $(\xavrd[_{t_j}],\cavrd[_{t_j}])$ two different formation profiles
  belonging to the same connected component\footnote{%
    Let $(\xavrd[_{t_{j-1}}],\cavrd[_{t_{j-1}}])\in\mathcal{S}$
    and $(\xavrd[_{t_{j}}],\cavrd[_{t_{j}}])\in\mathcal{S}$ refer to the formation profiles
    for $\bar{d}_{t_{j-1}}\z$ and $\bar{d}_{t_{j}}\z$, respectively. Let
    $r(z,\alpha)$ be given so that $r(z,0)=\bar{d}_{t_{j-1}}(z)$ and
    $r(z,1)=\bar{d}_{t_{j}}(z)$, e.g., $r(z,\alpha)=\bar{d}_{t_{j-1}}(z) +
    \alpha[\bar{d}_{t_{j}}(z)-\bar{d}_{t_{j-1}}(z)]$.
    It can be shown that $(\xavrd[_{t_{j-1}}],\cavrd[_{t_{j-1}}])$ and
    $(\xavrd[_{t_{j}}],\cavrd[_{t_{j}}])$ belong to the same connected
    component of $\mathcal{S}$, if the solution of
    \eqref{eq:sec2:problemStdyst} with $\bar{d}(z)$ replaced by
    $r(z,\alpha)$ is defined on $[0,\ell]$ for any $\alpha\in[0,1]$
    (see \cite{coron_trelat:04} for a related setting in the context of
    steady state controllability).
  } of $\mathcal{S}$. The
  transition from $(\xavrd[_{t_{j-1}}],\cavrd[_{t_{j-1}}])$ to
  $(\xavrd[_{t_{j}}],\cavrd[_{t_{j}}])$ in the finite time interval $t\in[t_{j-1},t_j]$ is
  achieved, if inputs $\uu[_0]$, $\uu[_\ell]$, $\vv[_0]$,
  $\vv[_\ell]$ exist for $t\in[t_{j-1},t_j]$ so that starting from 
  $(x(z,t_{j-1}),c(z,t_{j-1}))=(\xavrd[_{t_{j-1}}],\cavrd[_{t_{j-1}}])$ the
  solution $(x(z,t_{j}),c(z,t_{j}))=(\xavrd[_{t_{j}}],\cavrd[_{t_{j}}])$ is
  obtained. 
\end{enumerate}
\begin{rem}\label{rem:sec2:state:c}
  The inclusion of the second state $c\zt$ into the problem
  formulation extends the possible set of formation
  profiles. To illustrate this consider the computation of the steady state (only
  $\bar{c}$-contribution) according to \eqref{eq:sec2:problemStdyst}
  for (i) $\bar{d}_0\z=0$, i.e., $\cavr = \scavr_{0} +
  z(\scavr_{\ell}-\scavr_{0})$ and (ii) $\bar{d}_0\z=\pi^2$,
  $\scavr_{0}=\scavr_{\ell}=0$, i.e., $\cavr = k \sin(\pi z)$, $k\ne
  0$. Since there are obviously smooth functions $d\zt$ with $d(z,t_0)=0$,
  $d(z,t_1)=\pi^2$ for some $t_1>t_0$, locally non-analytic at
  $t\in\{t_0,t_1\}$ this example confirms that the $c$-dynamics can be
  controlled by $d\zt$ and the boundary inputs $\vv[_0]$ and
  $\vv[_\ell]$ to connect different families of steady states. The
  explicit constructive solution of this trajectory planning problem
  is presented in Section \ref{sec:flatnessFF}. 
\end{rem}

\begin{figure}[tb]
  \subfigure[][Circle formation \label{fig:sec2:stdyStCircle}]
  {\includegraphics{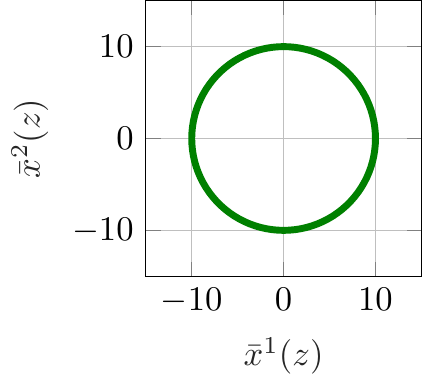}}
  \subfigure[][Gull-like formation \label{fig:sec2:stdyStGull}]
  {\includegraphics{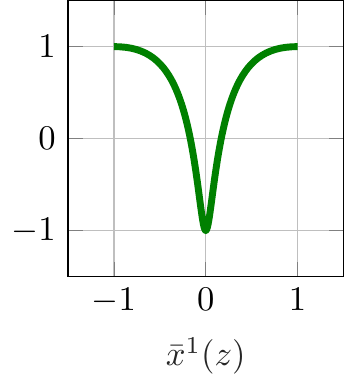}}
  \caption{Formation profiles from the overlay of solutions to
    \eqref{eq:sec2:problemStdyst} for parameter values according to
    Tab. \ref{tbl:sec2:ucStdyStExpl}.}
  \label{fig:sec:stdySts}
\end{figure}

While in general a numerical solution of \eqref{eq:sec2:problemStdyst}
is required, analytic expressions can be determined for special
cases. For $\bar{d}\z=0$ \eqref{eq:sec2:problemStdyst} yields
\begin{align}
  \begin{split}
    &a\spddz{\xavr} + \Big(\scavr_{0}+(\scavr_{\ell}-\scavr_{0})\frac{z}{\ell}\Big) \xavr = 0,\quad z\in(0,\ell)\\
    &\sxavr(0) = \sxavr_{0},~\sxavr(\ell)=\sxavr_{\ell},
  \end{split}
\end{align}
whose solution, without imposing additional conditions on the
coefficients, can be determined by means of Airy functions. Let in
addition $\scavr_{0}=\scavr_{\ell}$, then steady state formation
profiles $ \xavr $ can be written as 
\begin{align}\label{eq:sec2:stdyStAnsatz1}
  \xavr = \ksymb_1\exp{(\ssymb_1\zsymb)} + \ksymb_2\exp{(\ssymb_2\zsymb)}\,.
\end{align}
For $a>0$ three scenarios are possible: (i) If $\scavr_0<0$, then the
solution is \eqref{eq:sec2:stdyStAnsatz1} with
$\{s_1,s_2,k_1,k_2\}\in\mathbb{R}$; (ii) if $\scavr_0 > 0 $, then the solution reads
\begin{align}\label{eq:sec2:stdyStAnsatzCS}
  \xavr = k'\cos{(\theta\zsymb)} + k''\sin{(\theta\zsymb)}\,\quad k',~k''\in\mathbb{R};
\end{align}
and (iii) if $\scavr_0=0$ one obtains $\xavr = k_1 + k_2\zsymb$.
The explicit computation of the coefficients $ \ksymb_1 $ and $
\ksymb_2 $ relies on the values $\sxavr_0$ and $\sxavr_{\ell}$ of the
leader agents in \eqref{eq:sec2:problemStdyst}. 
Particular examples are shown in \figref{fig:sec:stdySts} when solving
the boundary value problem as described before individually for the
$\xsymb^{1}$- and the $\xsymb^{2}$-direction given the parameters of
Tab.~\ref{tbl:sec2:ucStdyStExpl}. In general the overlay
of solutions \eqref{eq:sec2:stdyStAnsatz1} in the two 
dimensional plane generates shapes of the well-known Lissajous curves.
Note that for the circle formation the parameter configuration for the
$\sx^2$-coordinate allows an arbitrary setting for $ k'' $ but with $
k'=0 $ in \eqref{eq:sec2:stdyStAnsatzCS}. Consequently, the steady
state solution is not uniquely determined but can be freely scaled in $ k'' $. 
\begin{table}[t]
  \caption{Parameters for the formation profiles in
    \figref{fig:sec:stdySts}.} 
  \centering
  \begin{tabular}{l r c c c c c}
    \toprule
    Profile$ \quad $ & Coord.  & $ a $  & $ \scavr_0,\,\scavr_\ell $ & $ \sxavr_{0} $ & $ \sxavr_{\ell} $\\
    \midrule
    \multirow{2}{*}{circle} & $ \sx^1: $ & $  1 $ & $ (2\pi/\ell)^2 $
                                                       &  $ 10 $ & $  10 $\\
                     & $ \sx^2: $ & $  1 $ & $ (2\pi/\ell)^2 $ & $  0 $ & $  0 $\\
    \multirow{2}{*}{gull-like}  & $ \sx^1: $ & $  1 $ & $ -(7/\ell)^2$ 
                                                       & $ -1 $ & $ 1 $\\
                     & $ \sx^2: $ & $  1 $ & $ (2\pi/\ell)^2$ & $  1 $ & $ 1 $\\
    \bottomrule
  \end{tabular}
  \label{tbl:sec2:ucStdyStExpl}
\end{table}
%


%
\section{Trajectory planning for agent continuum}\label{sec:flatnessFF}
%
Trajectory planning refers to the determination of the input
trajectories so that the system state or output follows a certain
predefined path. This problem is subsequently solved by
exploiting the flatness property of multi-agent continuum model
\eqref{eq:sec2:problem}.
%
\subsection{Formal state and input
  parametrisation}\label{sec:sec3:frmlParamtrstn}
%
To differentially parametrise the system state $\bs{\eta}\zt=[\x,\cc]^T$ and the boundary controls
$\bs{\kappa}_0\t=[\uu[_0],\vv[_0]]^T$ at $z=0$ and
$\bs{\kappa}_\ell\t=[\uu[_\ell],\vv[_\ell]]^T$ at $z=\ell$ formal 
integration as proposed in
\cite{meurer_trajectory_2009,meurer_control_2013} is extended to the
multi-input case with inputs on opposite boundaries of the domain.
Let
\begin{align*}
  \bs{f}(\bs{\eta},\zsymb,\tsymb) =
  \begin{bmatrix}
    \frac{1}{a}\big(\spdt{\x} - \cc\x\big)\\
    \frac{1}{b}\big(\spdt{\cc} - d\zt\cc\big)
  \end{bmatrix}
\end{align*}
and solve \eqref{eq:sec2:problemPDE:x}, \eqref{eq:sec2:problemPDE:c} for
$[\spddz{\x},\spddz{\cc}]^T$. Integrating the resulting
expression twice in $z$ yields
\begin{multline}
  \label{eq:sec3:frmlPrmtrstn1}
  \bs{\eta}\zt
  =
  \bs{\eta}(\xi,t)
  +
  (z-\xi)
  \spdz{\bs{\eta}\zt}\vert_{z=\xi}\\
  +
  \int^{z}_{\xi}
  \int^{\chi}_{\xi}
  \bs{f}(\bs{\eta},\sigma,\tsymb)
  \dx{\sigma}\dx{\chi}.
\end{multline}
for arbitrary but fixed $\xi\in[0,\ell]$. As a result
\begin{align}\label{eq:sec3:fltOutp}
  \bs{y}_1\t=\bs{\eta}(\xi,t),\quad
  \bs{y}_2\t=\spdz{\bs{\eta}\zt}\vert_{z=\xi}
\end{align}
serve as degrees-of-freedom. This enables us to implicitly express
$\bs{\eta}\zt$ and thus the boundary inputs
$(\bs{\kappa}_0\t,\bs{\kappa}_\ell\t)$ in terms of $(\bs{y}_1\t,\bs{y}_2\t)$
according to
\begin{subequations}
  \begin{align}
    \bs{\eta}\zt
    & =
      \bs{y}_1\t
      +
      (z-\xi)\bs{y}_2\t\nonumber\\
    &\phantom{=}
    +
    \int^{z}_{\xi}
    \int^{\chi}_{\xi}
    \bs{f}(\bs{\eta},\sigma,\tsymb)
      \dx{\sigma}\dx{\chi}\label{eq:sec3:statePrmtrstn}\\
    \bs{\kappa}_0\t &= \spdt{\bs{\eta}}(0,t),~
    \bs{\kappa}_\ell\t = \spdt{\bs{\eta}}(\ell,t)\label{eq:sec3:statePrmtrstn:inputs}
  \end{align}
\end{subequations}
An explicit expression can be obtained either by iteration or
successive approximation. For the latter consider the functional series
\begin{align}
  \label{eq:sec3:inftySeries}
  \bs{\eta}\zt = \sum_{n=0}^{\infty} \bs{\eta}_n\zt,
\end{align}
whose substitution into \eqref{eq:sec3:statePrmtrstn} motivates the
computational rule
\begin{align}\label{eq:sec3:seriesCeoff}
\begin{split}
  \bs{\eta}_0\zt &= \bs{y}_1\t + (z-\xi)\bs{y}_2\t\\
  \bs{\eta}_n\zt &= \int^{z}_{\xi}
    \int_{\xi}^{\chi}
    \bs{f}(\bs{\eta}_{n-1},\sigma,\tsymb)
      \dx{\sigma}\dx{\chi},\quad n\geq 1.
\end{split}
\end{align}
In other words $(\bs{y}_1\t,\bs{y}_2\t)$ defined in
\eqref{eq:sec3:fltOutp} for arbitrary $\xi\in[0,\ell]$ can be
considered a flat output for the multi-agent continuum model
\eqref{eq:sec2:problem}. The explicit evaluation of
\eqref{eq:sec3:inftySeries}, \eqref{eq:sec3:seriesCeoff} thereby
relies on the convergence of the obtained expressions, which, as is
shown below, reduces to a problem of trajectory assignment for the
flat output. 
%
\subsection{Convergence analysis}\label{sec:sec3:cnvrgcAnlys}
%
For the convergence analysis the notion of a Gevrey class function is
required \citep{rodino_gevrey_1993}.
\begin{defn}[Gevrey class functions]\label{def:app:gevreyClass}
  The function $y\t$ is in $ G_{\alpha}^{D}(\R)$, the Gevrey class of
  order $\alpha$, if $y\in C^\infty(\R)$ and $\exists D\in(0,\infty)$ so that $\sup_{t\in\R}\abs{\spdt[k]{y\t}} \leq
  D^{k+1}(k!)^\alpha$ holds true for all $k\in\N\cup\{0\}$. 
\end{defn}
\begin{defn}
  Let $G_\alpha^D(\mathbb{R};X)$ denote the class of 
  $X$-valued Gevrey class functions $G_{\alpha}^{D}(\R)$ of order
  $\alpha$ and let $\Omega_1\subset \R$, $\Omega_2\subset\R$. By
  $CG^{k,\alpha}(\Omega_1,\Omega_2;\R)$ we denote the class of
  functions $f:\Omega_1\times\Omega_2\to\R$ such that $f(\cdot,t)\in
  C^k(\Omega_1;\R)$ for every fixed $t\in\Omega_2$ and
  $f(z,\cdot)\in G_\alpha^D(\Omega_2;\R)$ for every fixed
  $z\in\Omega_1$. 
\end{defn}
The main convergence result reads as follows.
\begin{thm}\label{thm:sec3:convergence}
  Let $d\in CG^{0,\alpha}([0,\ell],\R;\R)$ and let
  $\bs{y}_1,\,\bs{y}_2\in (G_{\alpha}^{D}(\R))^{2}$ with
  $\alpha\leq 2$. Then $\bs{\eta}\zt$ determined from the series
  \eqref{eq:sec3:inftySeries} with coefficients
  \eqref{eq:sec3:seriesCeoff} fulfills $\bs{\eta}\in
  (CG^{2,\alpha}([0,\ell],\R;\R))^{2}$ and the series
  \eqref{eq:sec3:inftySeries} converges absolutely and uniformly for
  all $z\in[0,\ell]$ if $\alpha\in[1,2)$. 
\end{thm}
The proof of this result follows in principle from the
analysis\footnote{The fact that the same constants $D$ are used does not
  restrict generality since one may take $D=\max\{D_d,D_y\}$ with $D_d$
  and $D_y$ the individual Gevrey class constants for $d\zt$ and
  $\bs{y}\t$. The same holds true for $\alpha$ which is considered as
  $\alpha=\max\{\alpha_d,\alpha_y\}$.} in \cite{meurer_trajectory_2009} but with the modification
that the flat output is located at some fixed but arbitrary in-domain\footnote{Note that in-domain flat outputs have been addressed
already in \cite{rudolph_etal:05,meurer_finite-time_2011} taking into account power
series. The approach considered here generalizes these results since
$d\zt$ is not assumed to allow a power series expansion in $z$.} position
$\xi\in[0,\ell]$.
\begin{pf}
  For the convergence analysis the cascaded structure of the PDEs
  \eqref{eq:sec2:problem} is exploited by first analyzing the
  differential parametrization of $\eta_2\zt=c\zt$. 
  Taking into account the assumptions on $\bs{y}_1\t$, $\bs{y}_2\t$
  and $d\zt$ and the recursion \eqref{eq:sec3:seriesCeoff} it can be
  rather straightforwardly verified by induction that the
  $\N\cup\{0\}\ni l$-th time derivative of $\eta_{2,n}\zt=c_n\zt$
  fulfills
  \begin{align}
    \label{eq:sec3:conv:proof:c}
    \sup_{t\in\R} \vert\partial_t^{l}c_n\zt\vert \leq
    D^{l+n+1}(l+n)!^{\alpha} M_n g_n(z,\xi)
  \end{align}
  with $M_n = \frac{1}{b^{n}}\prod_{j=1}^{n}(1+\frac{1}{j^{\alpha}})$,
  $M_0=1$ and $g_n(z,\xi) = {\vert z-\xi\vert^{2n}}/{(2n)!} +
    {\vert z-\xi\vert^{2n+1}}/{(2n+1)!}$.
  Observing $b^{n}M_n =
  \prod_{j=1}^{n}(1+j^{\alpha})/\prod_{j=1}^{n}j^{\alpha}
  \leq (n+1)^{\alpha}$ the estimate
  \eqref{eq:sec3:conv:proof:c} for $l=0$ implies
  \begin{align*}
    \sup_{t\in\R} \vert c_n\zt\vert
    &\leq
      D^{n+1}(n)!^{\alpha} M_n g_n(z,\xi)\\
    &\leq D\bigg(\frac{D}{b}\bigg)^{n} (n+1)!^{\alpha} g_n(z,\xi)
  \end{align*}
  In view of \eqref{eq:sec3:inftySeries} and $g_n(z,\xi)
  =\vert z-\xi\vert ^{2n}/(2n)! \times (1+\vert z-\xi\vert)/(2n+1)) \leq
  (1+\ell)(z-\xi)^{2n}/(2n)!$ this yields the upper power series
  estimate on the functional series 
  \begin{align*}
    \vert c\zt\vert
    &\leq
      D \sum_{n=0}^{\infty}
      \bigg(\frac{D}{b}\bigg)^{n} (n+1)!^{\alpha} g_n(z,\xi) \\
    &\leq
      D(1+\ell) \sum_{n=0}^{\infty}
      \frac{(n+1)!^{\alpha}}{(2n)!}
      \kappa^{n} = \sum_{n=0}^{\infty} \beta_n\kappa^{n}
  \end{align*}
  with $\kappa=D\vert z-\xi\vert^{2}/b$. Absolute and uniform convergence with
  infinite radius of convergence for $\alpha\in[1,2)$ hence 
  follows from the Cauchy-Hadamard theorem applied to the coefficient
  $\beta_n$.

  By proceeding similarly with $c_n\zt$ replaced by
  $\partial_t^{l}c_n\zt$ and the construction in terms of formal
  integration it can be deduced that $c\zt$ obtained from
  \eqref{eq:sec3:inftySeries}, \eqref{eq:sec3:seriesCeoff} fulfills
  $c\in CG^{2,\alpha}([0,\ell],\R;\R)$. This result directly
  implies that the convergence analysis for $\eta_1\zt=x\zt$ follows
  exactly the lines above for $c\zt$, which proves the claim.  \qed
\end{pf}
%
%
\subsection{Trajectory assignment}
%
Based on the flatness analysis
above desired trajectories for the flat outputs $\bs{y}_1\t$ and
$\bs{y}_2\t$ can be assigned independently to achieve prescribed
finite time transitions between formation profiles. According to
Section \ref{sec:sec2:frmtnProfiles} these are completely
determined by solving the boundary-value problem
\eqref{eq:sec2:problemStdyst}. Let $(\sxavr_{0}^{\ast},\sxavr_{\ell}^{\ast})$ and
$(\scavr_{0}^{\ast},\scavr_{\ell}^{\ast})$ denote the desired boundary
values \eqref{eq:sec2:problemStdyst:bc} of the formation
$(\xavrd,\cavrd)$. With \eqref{eq:sec2:stdyStSol} and
\eqref{eq:sec3:fltOutp} the resulting formation profile can be
translated into steady state values of the flat outputs according to 
\begin{align}
  \label{eq:sec3:FltOutpConst}
  \bs{y}_1^{\ast}
  =
    \begin{bmatrix}
      \sxavr^{\ast}(\xi;\sxavr_{0}^{\ast},\sxavr_{\ell}^{\ast})\\
      \scavr^{\ast}(\xi;\scavr_{0}^{\ast},\scavr_{\ell}^{\ast})
    \end{bmatrix},
  \quad
   \bs{y}_2^{\ast}
  =
    \begin{bmatrix}
      \partial_z\sxavr^{\ast}(\xi;\sxavr_{0}^{\ast},\sxavr_{\ell}^{\ast})\\
      \partial_z\scavr^{\ast}(\xi;\scavr_{0}^{\ast},\scavr_{\ell}^{\ast})
    \end{bmatrix}.
\end{align}
By changing $(\sxavr_{0}^{\ast},\sxavr_{\ell}^{\ast})$ and
$(\scavr_{0}^{\ast},\scavr_{\ell}^{\ast})$ different formation
profiles are obtained, which can be connected by properly assigning
the temporal transition path for the flat output. To illustrate this
let $\bs{y}_{1,0}^{\ast}$, $\bs{y}_{2,0}^{\ast}$ and
$\bs{y}_{1,\tau}^{\ast}$, $\bs{y}_{2,\tau}^{\ast}$ denote steady state
values determined from \eqref{eq:sec3:FltOutpConst} corresponding to
two formation profiles $(\xavrd[_{t_0}],\cavrd[_{t_0}])$ and
$(\xavrd[_{t_1}],\cavrd[_{t_1}])$ to be attained at times $t=t_0$ and
$t=t_1=t_0+\tau$, respectively. The transition between these two profiles
within the finite time interval $t\in[t_0,t_0+\tau]$ can be realized
by assigning 
\begin{align}\label{eq:sec3:FltOutp:j}
  \yvdes[_j] &= \bs{y}_{j,0}^{\ast} +
  \big(\bs{y}_{j,\tau}^{\ast}-\bs{y}_{j,0}^{\ast}\big)
  \Phi_{\tau}(t-t_0) 
\end{align}
for $j\in\{1,2\}$. Herein, $\Phi_\tau(\cdot)$ has to be a Gevrey class
function according to Def. \ref{def:app:gevreyClass} being locally
non-analytic at $t=t_0$ and $t=t_0+\tau$, i.e., $\Phi_\tau(0)=0$,
$\Phi_\tau(\tau)=1$ with
$\partial_t^{l}\Phi_\tau\vert_{t\in\{0,\tau\}}=0$ for $l\in\N$.
The latter requires a Gevrey order $\alpha>1$ with $\alpha<2$ being
imposed from Thm. \ref{thm:sec3:convergence}. Examples for functions $\Phi_\tau(\cdot)$ are provided, e.g., in 
\cite{rodino_gevrey_1993,laroche:00}. 

Moreover, given an arbitrary formation profile $\xavrdd[\z]$, which does
not fulfill \eqref{eq:sec2:problemStdyst} the presented approach can
be extended to approximately obtain the desired profile. For this, the
static optimization problem is formulated
\begin{align}
  \begin{split}
  &\min_{\scavr_{0},\scavr_{\ell},\bar{d}(z)}
    J\big[\xavr-\xavrdd[\z]\big]\\
  &\text{s.t.}~\xavr~\text{fulfills \eqref{eq:sec2:problemStdyst}
    for}~\sxavr_{0}=\xavrdd[(0)],\,\sxavr_{\ell}=\xavrdd[(\ell)].
  \end{split}
\end{align}
Herein, $J[\cdot]$ is a positive definite functional to be chosen
suitably depending on the problem to minimize the difference between
the steady state $\xavr$ and the desired formation profile $\xavrdd[\z]$.   
%
\subsection{Feedforward control}
%
Given the desired flat output trajectories the corresponding
feedforward control signals follow from \eqref{eq:sec3:statePrmtrstn:inputs} with $\bs{\eta}\zt$ replaced by
$\bs{\eta}^{\ast}\zt$ computed in terms of the series
\eqref{eq:sec3:inftySeries} with recursively evaluated coefficients
\eqref{eq:sec3:seriesCeoff} in terms of $\bs{y}_1^{\ast}\t$,
$\bs{y}_2^{\ast}\t$. 
This yields
\begin{align}\label{eq:sec3:fdfdterms}
  \bs{\kappa}_0^{\ast}\t = \spdt{\bs{\eta}^{\ast}}(0,t),\quad
  \bs{\kappa}_\ell^{\ast}\t = \spdt{\bs{\eta}^{\ast}}(\ell,t).
\end{align}
%


\section{Observer-based tracking control}
\label{sec:trckngCtrl}
%
Since formation profiles may correspond also to unstable steady states
of the PDE a stabilizing feedback control is required. In view of
motion planning and the resulting feedforward control subsequently the
spatial-temporal tracking error is stabilized using
a backstepping approach involving a distributed parameter state
observer. This results in a so-called two-degrees-of-freedom (2DOF)
control approach with the desired motion induced by the feedforward
control and the stabilization provided by the feedback control.
%
\subsection{Stabilisation of tracking error dynamics}\label{sec:bckstStabCtrl}
%
The state $c\zt$ is used to distribute information
to the PDEs \eqref{eq:sec2:problemPDE:x} governing the agent position
$\x$.
\begin{assum}
  \label{assump:stability:csys}
  The solution to \eqref{eq:sec2:problemPDE:c},
  \eqref{eq:sec2:c:cont:leader} with initial state \eqref{eq:sec2:problemIC} fulfills $c^{i}\zt\in
  CG^{2,\alpha}([0,\ell],\R;\R)$, $\alpha\leq 2$. 
\end{assum}
This assumption can be fulfilled in a straightforward way by a
proper choice of $d^i\zt$, see also the main convergence result in
Theorem \ref{thm:sec3:convergence}, and implies that $c^{i}\zt$ is
bounded.  
In view of Assumption \ref{assump:stability:csys} and the
cascaded structure consisting of 
\eqref{eq:sec2:problemPDE:x} and \eqref{eq:sec2:problemPDE:c} the
sub-dynamics for $c\zt$ is subsequently assumed to be only controlled
by the feedforward control $(\vv[_0^\ast],\vv[_\ell^\ast])$. To
  emphasize this fact $c^\ast\zt$ is written subsequently when referring to
  this solution. Contrary
the sub-dynamics for $\x$ is controlled using a combined
feedforward-feedback strategy.
Since flatness-based motion planning by construction fulfills the PDE
\eqref{eq:sec2:problemPDE:x} with $(\x,\un,\ul)$ replaced by $(\xdes,\ffn,\ffl)$
the tracking error dynamics in the error state $\xerrc=\x-\xdes$ reads
\begin{align}
  \label{eq:sec4:errSysCtrl}
  \begin{split}
    \spdt{\xerrc}  &= a\spddz{\xerrc} + c^\ast\zt\xerrc\\
    \sbcn{\sxerrc} &= \un - \ffn = \fbn\\
    \sbcl{\sxerrc} &= \ul - \ffl = \fbl.
  \end{split}
\end{align}
Herein, $\fbn$ and $\fbl$ are used to establish state feedback
control. For this backstepping is used by introducing the invertible
time-varying Volterra integral transformation
\begin{align}\label{eq:sec4:bstepTrnsfCtrl}
  \vtsymb\zt = \xerrc - \integral{k\zst \sxerrc\st}{s}{0}{z},
\end{align}
with the integral kernel $k\zst$ defined on $\zst \in
\mathcal{T}_k(\ell):=\ofB{\zst \in \R^2 \times \Rtn \mid
  s\in[0,\ell], z\in\ofb{s,\ell}}$ to invertibly map
\eqref{eq:sec4:errSysCtrl} into the target system
\begin{align}
  \label{eq:sec4:tarSysCtrl}
  \begin{split}
    \sdcr[{&}]{\vtsymb\zt}{a}{0}{0}-\mu\t\vtsymb\zt\\
    &\phantom{=}\,- a\spd{\ksymb\znt}{\ssymb} \vtsymb\nt,~z\in(0,\ell),\,t>t_0\\
    \spdt\vtsymb\nt &= -\mu\t \vtsymb\nt~
    \spdt\vtsymb\lt = -\mu\t \vtsymb\lt\\
    \vtsymb(z,t_0) &= \vtsymb_0\z,~z\in[0,\ell].
  \end{split}
\end{align}
with the time-varying design parameter $ \mu\t$, see also 
\cite{frihauf_leader-enabled_2011} for a related but time-invariant
case. 
Differentiating \eqref{eq:sec4:bstepTrnsfCtrl} once with respect to
$t$ and twice with respect to $z$ followed by the substitution of
\eqref{eq:sec4:tarSysCtrl} leads, after some interim but straightforward
calculations (see, e.g., \cite{meurer_control_2013}), to the
well-known kernel equations 
\begin{align}\label{eq:sec4:intKernPDECtrl}
  \begin{split}
    \spdt\ksymb\zst &=a\spddz{\ksymb}\zst - a\spdd{\ksymb}{s}\zst\\
    &\qquad-(c^\ast\st+\mu\t)\ksymb\zst\\
    \ksymb\zzt &= -\frac{1}{2a}\intsntoz{\ofB{c^\ast\st+\mu\t}}\\
    \ksymb\znt &= 0.
  \end{split}
\end{align}
For the determination of the solution $\ksymb\zst$ of
\eqref{eq:sec4:intKernPDECtrl} using either formal integration and
successive approximation or a suitable numerical scheme the reader is
referred to, e.g.,
\cite{meurer_tracking_2009,jadachowski_efficient_2012}. With
  Assumption \ref{assump:stability:csys} it can be shown that 
$\ksymb\zst$ is a strong solution to \eqref{eq:sec4:intKernPDECtrl}
with $\ksymb\in
  CG^{2,\alpha}(\Gamma,\R;\R)$, $\Gamma=\{(z,s)\in \R^2 \mid
  s\in[0,\ell], z\in[s,\ell]\}$
\citep{vazquez_etal:08,meurer_tracking_2009}. 

The state feedback controllers $ \fbn $ and $ \fbl $ follow by
evaluating \eqref{eq:sec4:bstepTrnsfCtrl} and its time derivative at
the boundaries together with \eqref{eq:sec4:errSysCtrl} and
\eqref{eq:sec4:tarSysCtrl}. With this, the controller at $z=0$ reads 
\begin{subequations}
  \begin{align}
    \label{eq:sec4:anchorLaw}
    \fbn= -\mu\t \sxerrc\nt.
  \end{align}
  The evaluation at $ z=\ell $ yields a more complex expression
  \begin{align}
    \label{eq:sec4:leaderLaw}
    \begin{split}
      \fbl &= -\ofb{\mu\t+ a\spd{k\llt}{\ssymb}}\sxerrc\lt \\
      +& \integral{\!\!\!\!\ksymb_{I}\lst\sxerrc\st}{s}{0}{\ell} + a k\llt\spdz{\sxerrc\lt}\\
      +&\,a \spd{k\lnt}{\ssymb}\sxerrc\nt
    \end{split}
  \end{align}
\end{subequations}
with $\ksymb_{I}\lst = \ofb{\mu\t+\sc\st}k\lst +
\spdt{k\lst} + \alpha \spd[2]{k\lst}{\ssymb}$. This expression results
from the evaluation of the boundary condition for $z=\ell$ in
\eqref{eq:sec4:tarSysCtrl} taking into account
\eqref{eq:sec4:errSysCtrl}, \eqref{eq:sec4:bstepTrnsfCtrl} and using
partial integration twice. The existence of the derivatives
$\spdt{k\lst}$, $\spd{k\lst}{\ssymb}$ and $\spd[2]{k\lst}{\ssymb}$ in
\eqref{eq:sec4:leaderLaw} follows from $\ksymb\zst$ being a strong
solution having Gevrey properties in $t$. 
%
\subsection{Closed-loop stability analysis}\label{sec:fbCtrlStabAnlys}
%
Subsequently well-posedness and stability of the target dynamics
\eqref{eq:sec4:tarSysCtrl} are analysed by considering the governing
equations in the space $X=H^1(0,\ell)$ equipped with the norm
$\|h\|_X=\sqrt{\langle h,h \rangle_X}$ induced by the inner
product $\langle h_1,h_2
\rangle_X=h_1(0)h_2(0)+h_1(\ell)h_2(\ell)+\langle \spdz h_1,\spdz h_2\rangle_{L^2}$
for $h,\,h_1,\,h_2\in X$. 
It is also referred, e.g., to \cite{liang_etal:03} for a general
Banach space analysis in the non-autonomous case. 

By (i) introducing the transformation
$\vtsymb\zt=\exp(-\int_{t_0}^{t}\mu(\tau)\mathrm{d}\tau)\mathfrak{y}\zt$
to remove the terms involving $\mu(t)$ from
\eqref{eq:sec4:tarSysCtrl} followed by (ii) homogenizing the boundary
conditions using $\mathfrak{x}\zt = \mathfrak{y}\zt +
b_0(z)\vtsymb_0(0) + b_{\ell}(z)\vtsymb_0(\ell)$ with
$b_0(z)=z/\ell-1$, $b_\ell(z)=-z/\ell$ one obtains
$\sdcr[]{\mathfrak{x}\zt}{a}{0}{0}- a\spd{\ksymb\znt}{\ssymb}
\vtsymb_0(0),~z\in(0,\ell),~t>t_0$ subject to
$\mathfrak{x}(0,t)=\mathfrak{x}(\ell,t)=0$, $\mathfrak{x}(z,t_0)=\mathfrak{x}_0(z)=\vtsymb_0(z) +
b_0(z)\vtsymb_0(0) + b_{\ell}(z)\vtsymb_0(\ell)$. The solution of
the resulting inhomogeneous PDE can be determined using separation
of variables and Fourier expansion. After reverting
steps (ii) and (i) this yields the solution
\begin{multline*}
  \vtsymb\zt = e^{-\int_{t_0}^{t}\mu(\tau)\mathrm{d}\tau}\bigg(
  S(t)\vtsymb_0(z) - b_0(z)\vtsymb_0(0)\\ -
  b_{\ell}(z)\vtsymb_{0}(\ell)- \int_{t_0}^{t}
  S(t-\tau)a\spd{\ksymb\znt}{\ssymb} 
  \vtsymb_0(0)\mathrm{d}\tau  \bigg),
\end{multline*}
where 
\begin{align*}
  S(t)h = \sum_{k=1}^{\infty} e^{\lambda_k(t-t_0)}\langle
  h,\phi_k(z)\rangle_X \phi_k(z),\quad h\in X
\end{align*}
with $\lambda_k=-a(k\pi/\ell)^2$, $\phi_k(z) = A_k \sin(k\pi
z/\ell)$, $A_k = \sqrt{2\ell}/(k\pi)$ for $k\in\N$ is a $C_0$-semigroup on $X$. 
By applying the Gram-Schmidt orthogonalisation
  procedure the functions $\phi_{-1}(z)={1}/{\sqrt{2}}$, 
  $\phi_{0}(z)=({z}/{\ell}-{1}/{2})/\sqrt{{1}/{\ell}+{1}/{2}}$ can be
  determined from $b_0(z)$, $b_\ell(z)$ so that
  $\mathcal{B}=\{\phi_{-1}(z),\phi_{0}(z),\phi_1(z),\phi_2(z),\ldots,\phi_k(z),\ldots\}$
  is an orthonormal set, i.e., $\langle \phi_i,\phi_j
  \rangle_X=\delta_{i,j}$ for $i,j\in\N\cup\{-1,0\}$. Since 
  $b_0(z)=-{1}/{\sqrt{2}}\phi_{-1}(z)+\sqrt{{1}/{\ell}+{1}/{2}}\phi_0(z)$ and
  $b_\ell(z)=-{1}/{\sqrt{2}}\phi_{-1}(z)-\sqrt{{1}/{\ell}+{1}/{2}}\phi_0(z)$
  it follows that $\langle b_j(z),\phi_k(z)\rangle_X=0~\forall k\in\N$,
  $j\in\{0,\ell\}$, which is used to simplify $S(t)\mathfrak{x}_0(z)$
  when solving for $\vtsymb\zt$.
  Moreover it can
  be shown that $\langle h,\phi_j \rangle_X=0$, $j\in\N\cup\{-1,0\}$ implies
  $h=0$. Hence $\mathcal{B}$ is maximal and as a consequence is a
  complete orthonormal basis of $X$, see, e.g. \cite[Prop. 5.36,
  5.38]{kubrusly:11}.
For the homogeneous problem with $a\spd{\ksymb\znt}{\ssymb}=0$
the orthonormality property enables us to show in a straightforward way that 
$\|\vtsymb\|_X\leq\exp(-\int_{t_0}^{t}\mu(\tau)\mathrm{d}\tau)\|\vtsymb_0\|_X$. This 
confirms the continuous dependence of the solution on the initial
state and hence well-posedness in the sense of
Hadamard. Depending on the regularity of the inhomogeneity $\spd{\ksymb\znt}{\ssymb}$ classical
or mild solutions can be defined. In fact
$\spd{\ksymb\zst}{\ssymb}\in
CG^{1,\alpha}(\Gamma,\R;\R)$ so that for any
$\vtsymb_0\in X$ one has $\vtsymb\in C([t_0,\infty);X)\cap
C^1((t_0,\infty);X)$ with $\vtsymb\zt$ fulfilling
\eqref{eq:sec4:tarSysCtrl} pointwise.
Furthermore the analysis supports the following stability
result, which generalizes the approach in
\cite{frihauf_leader-enabled_2011} to the considered time-varying setup. 
\begin{lem}\label{lem:sec4:stability:target}
  Let $0<\epsilon^-<\mu\t\leq\epsilon^+<\infty$ for all $t\geq
    t_0$. Then the zero equilibrium of the target dynamics
    \eqref{eq:sec4:tarSysCtrl} is exponentially stable in the norm
    $\|\cdot\|_X$, i.e., there exists $M>0$ so that the inequality
    holds true
    \begin{multline}
      \label{eq:lem:sec4:stability:target}
      \|\vtsymb\|_X\t
      \leq M
      e^{-\frac{1}{2}\int_{t_0}^{t}\mu(s)\mathrm{d}s}
      \|\vtsymb\|_X(t_0)\\
      \leq M
        e^{-\frac{\epsilon^-}{2}(t-t_0)}\|\vtsymb\|(t_0).
    \end{multline}
  \end{lem}
  \begin{pf}
    Consider the Lyapunov functional
    \begin{align}
      \label{eq:sec4:stab:proof:layp:fun}
      V(t) = \frac{1}{2}\big[p\vtsymb^2(0,t) +\vtsymb^2(\ell,t)\!+\!
      \|\spdz\vtsymb\|^2_{L^2}\t\big]
    \end{align}
    with $p\geq 0$ to be determined below. There exist positive constants
    $0<\beta^{-}<\beta^{+}$ so that
    \begin{align}
      \label{eq:sec4:stab:proof:layp:est}
      \beta^{-}\|\vtsymb\|_X^{2}\t\leq V(t)\leq \beta^{+}\|\vtsymb\|_X^{2}\t.
    \end{align}
    A possible choice is $\beta^{-}=\min\{1/2,p/2\}$ and $\beta^{+}=\max\{1,p/2\}$.  
    The rate of change of $V(t)$ along a solution of
    \eqref{eq:sec4:tarSysCtrl} results in\footnote{To simplify
      expressions the explicit dependency of the variables on $z$ and
      $t$ is omitted when clear from the context.}
    \begin{align*}
      \spdt {V} &= p\vtsymb(0)\spdt\vtsymb(0) +
                   \vtsymb(\ell)\spdt\vtsymb(\ell) +
                \int_{0}^{\ell}\spdt\spdz\vtsymb\spdz\vtsymb
                \mathrm{d}z.
    \end{align*}
    Interchanging $\spdt\spdz\vtsymb=\spdz\spdt\vtsymb$,
    integrating by parts and substituting \eqref{eq:sec4:tarSysCtrl}
    using $f(z,t):=a\spd{\ksymb\znt}{\ssymb}$ gives
    \begin{align*}
      \spdt  {V} =&
                    \int_{0}^{\ell}\spdz[2]\vtsymb\big[a\spdz[2]\vtsymb-\mu\t\vtsymb-
                    f\vtsymb(0)\big]\mathrm{d}z\\
                  &-\mu\t\big[\spdz\vtsymb(\ell)\vtsymb(\ell)-\spdz\vtsymb(0)\vtsymb(0)+p\vtsymb^2(0)+\vtsymb^2(\ell)\big]\\
      =& -\mu\t \big[ p\vtsymb^2(0) + \vtsymb^2(\ell) + \|\spdz\vtsymb\|^2_{L^2}\big]\\
      & -a \|\spdz[2]\vtsymb\|^2_{L^2} + \int_{0}^{\ell}f\vtsymb(0)
         \spdz[2]\vtsymb \mathrm{d}z.
    \end{align*}
    Application of Cauchy-Schwarz and Young inequality to the last
    term, i.e.,
    $\int_{0}^{\ell}f\vtsymb(0)\spdz[2]\vtsymb \mathrm{d}z \leq
    \int_{0}^{\ell} \vert f\vert
    \vert\vtsymb(0)\vert\vert\spdz[2]\vtsymb\vert \mathrm{d}z\leq
    \frac{\rho}{2} \vtsymb^2(0)\|f\|^2_{L^2}+
    \frac{1}{2\rho} \|\spdz[2]\vtsymb\|^2_{L^2}$ for
    $\rho>0$, together with the boundedness of the kernel
    $K_s = \max_{t\geq t_0}\|f\|^2_{L^2} = \max_{t\geq
      t_0}\int_{0}^{\ell}(a\spd{\ksymb(z,0,t)}{\ssymb})^2\mathrm{d}z$
    implies 
    \begin{multline*}
      \spdt V \leq  -\bigg(\mu\t-\frac{K_s\rho}{2p}\bigg) p \vtsymb^2(0)
      -\mu\t \vtsymb^2(\ell)\\ - \mu\t \|\spdz\vtsymb\|^2_{L^2}- \bigg(a-\frac{1}{2\rho}\bigg)\|\spddz\vtsymb\|^2_{L^2}.
    \end{multline*}
    The inequalities $\mu\t-{K_s\rho}/{(2p)} \geq
    {\mu\t}/{2}$ and $a-{1}/{(2\rho)}\geq 0$
    are in view of the assumption
    $0<\epsilon^-<\mu\t\leq\epsilon^+<\infty$ for all $t\geq t_0$
    fulfilled, if $p\geq {\rho K_s}/\epsilon^{-}$ and
    $\rho\geq 1/(2a)$. Thus, one obtains $\dot{V}\t\leq -\mu\t V\t$. 
    Taking into account \eqref{eq:sec4:stab:proof:layp:est} the
    previous estimate implies \eqref{eq:lem:sec4:stability:target} with
    $M=\sqrt{\beta^{+}/\beta^{-}}$. \qed
  \end{pf}
  Note that the proof of Lemma \ref{lem:sec4:stability:target} can be performed
  identically for $p=1$ in $V(t)$ if the introduced constant $\rho$ can be
  bounded as ${1}/{(2a)}\leq \rho\leq {\epsilon^+}/{K_s}$.
  Lemma \ref{lem:sec4:stability:target} can be improved to
  verify pointwise exponential stability.
  \begin{cor}\label{cor:sec4:stability:target:sup}
    Let $0<\epsilon^-<\mu\t\leq\epsilon^+<\infty$ for all $t\geq
    t_0$. Then the zero equilibrium of the target dynamics
    \eqref{eq:sec4:tarSysCtrl} is exponentially stable in the $\sup$-norm
    $\|\cdot\|_\infty$, i.e., there exists $M>0$ so that
    \begin{multline}
      \label{eq:cor:sec4:stability:target}
      \sup_{z\in[0,\ell]}\vert\vtsymb\zt\vert
      \leq M 
      e^{-\frac{1}{2}\int_{t_0}^{t}\mu(s)\mathrm{d}s}\|\vtsymb\|_1(t_0)\\
      \leq M
      e^{-\frac{\epsilon^-}{2}(t-t_0)}\|\vtsymb\|_1(t_0)
    \end{multline}
    holds true with $\|h\|_1^2=h^2(0)+h^2(\ell)+\|h\|_{L^2}^2+\|\spdz 
    h\|_{L^2}^2$ for $h\z\in H^1(0,\ell)$. 
  \end{cor}
  \begin{pf}
    Taking into account the definition of the norm $\|\cdot\|_1$ there exist constants $0<\gamma^-<\gamma^+$ so that the
    Lyapunov functional $V\t$ introduced in
    \eqref{eq:sec4:stab:proof:layp:fun} can be bounded according to 
    \begin{align*}
      \gamma^{-} \|\vtsymb\|^2_1\t\leq V\t \leq \gamma^{+}
      \|\vtsymb\|^2_1\t. 
    \end{align*}
    The constants herein follow as $\gamma^{-} =
    {1}/{2}\min\{p,1-{r}/{(2\ell)},{r}/{(4\ell^2)},1-r\}$ and $\gamma^{+} = {1}/{2}\min\{1,p\}$
    with $0<r<\min\{1,2\ell\}$. While $\gamma^+$ can be directly
    deduced the determination of $\gamma^-$ requires to split the 
    term
    $\|\spdz\vtsymb\|^2_{L^2}=(1-r)\|\spdz\vtsymb\|^2_{L^2} + r\|\spdz\vtsymb\|^2_{L^2}$
    with $0<r<1$ in $V\t$ and to take into account the Poincar\'e
    inequality providing 
    $\|\spdz\vtsymb\|^2_{L^2}\geq
    1/(4\ell^2)\|\vtsymb\|^2_{L^2} -
    1/(2\ell)\vtsymb^2(\ell)$.\\[2ex]
    Noting that the analysis of $\dot{V}\t$ from the proof of Lemma
    \ref{lem:sec4:stability:target} carries over to the present case,
    i.e., $\dot{V}\t\leq -\mu\t V\t$ with $p$ and $\rho$ as before,
    one obtains using Agmon's and Young's inequality that
    \begin{multline*}
      \max_{z\in[0,\ell]}\vert\vtsymb\zt\vert^2  \leq \vtsymb^2(0) +
                                                  2\|\vtsymb\|\t\|\spdz\vtsymb\|\t\\
                                                 \leq \|\vtsymb\|^2_1\t \leq \frac{1}{\gamma^{-}}V\t
                                                \leq \frac{1}{\gamma^{-}}
                                                  e^{-\int_{t_0}^{t}\mu(s)\mathrm{d}s}
                                                  V(t_0)\\
                                                 \leq \frac{\gamma^{+}}{\gamma^{-}}
                                                  e^{-\int_{t_0}^{t}\mu(s)\mathrm{d}s} \|\vtsymb_0\|_1^2.
    \end{multline*}
    Substituting $M=\sqrt{\gamma^{+}/\gamma^{-}}$ verifies the claim. \qed
  \end{pf}
Proceeding similar to, e.g.,
\cite{meurer_tracking_2009,meurer_control_2013} one can by a direct
computation determine the inverse to \eqref{eq:sec4:bstepTrnsfCtrl}
given in the form  $ \xerrc = \vtsymb\zt  + \integral{g\zst
  \vtsymb\st}{\ssymb}{0}{\zsymb}$. Kernel equations for $g\zst$ can be
derived and it can be shown using straightforward arguments that the
differentiability properties of $\ksymb\zst$ carry over to the kernel
$g\zst$.
With Lemma \ref{lem:sec4:stability:target} it is a rather standard
procedure taking into account the boundedness of the kernel and the
inverse kernel as well as the Cauchy-Schwarz inequality to deduce the
stability of the closed-loop control system consisting of
\eqref{eq:sec4:errSysCtrl}, \eqref{eq:sec4:anchorLaw} and
\eqref{eq:sec4:leaderLaw} (see, e.g.,
\cite{frihauf_leader-enabled_2011,meurer_tracking_2009,meurer_control_2013}). In
particular there exist constants $C_0,\,C_1>0$ so that the following sequence
holds true 
\begin{multline}
  \|\sxerrc\|_X\t \leq C_0 \|\vtsymb\|_X\t \leq C_0
  e^{-\frac{1}{2}\int_{t_0}^{t}\mu(s)\mathrm{d}s}\|\vtsymb\|_X(t_0)\\
  \leq C_0 C_1
  e^{-\frac{1}{2}\int_{t_0}^{t}\mu(s)\mathrm{d}s} \|\sxerrc\|_X(t_0).
\end{multline}
Corollary \ref{cor:sec4:stability:target:sup} implies a
  similar result for $\|\sxerrc\|_1\t$.
%
\subsection{State observer design}\label{sec:observer}
%
The realization of the state feedback control composed of
\eqref{eq:sec4:anchorLaw} and \eqref{eq:sec4:leaderLaw} requires to
estimate the spatial-temporal evolution of $\sx\zt$ or
$\sxerrc\zt=\xest-\xdes$, respectively. Given \eqref{eq:sec2:problem}
the state observer is 
composed of a simulator and a correction part with the latter
injecting the considered output
\begin{align}
  \label{eq:sec4:output}
  \bs{o}\t=[\sx\nt,\sx\lt,\spdz{\sx\lt}]^T.
\end{align}
This results in 
\begin{align}\label{eq:sec4:obsvSys}
  \begin{split}
    \sdcr[{&}]{\xest}{a}{0}{c^\ast\zt}\\
    & \quad + L\zt(\sx\lt -\sxest\lt) \\
    & \quad + M\zt (\spdz{\sx\lt} - \spdz{\sxest\lt})\\
    \sbcn{\sxest} & = \un + l_0\t( \sx\nt - \sxest\nt )\\
    \sbce{\sxest} & = \ul + l_\ell\t( \sx\lt - \sxest\lt )\\
    \sxest\ztn & = \sxest_0\z,
  \end{split}
\end{align}
where $ \xest $ denotes the estimated state. The weights $ L\zt $, $
M\zt $, $l_0\t$, and $l_\ell\t$  are designed to ensure exponential
convergence of the observer error dynamics.
Introducing the observer error state $ \xerro=\x-\xest $ and taking
into account \eqref{eq:sec2:problem}, \eqref{eq:sec4:obsvSys} the
observer error dynamics is described by
\begin{align}\label{eq:sec4:obsvErrSys}
  \begin{split}
    \sdcr[{&}]{\xerro}{a}{0}{c^\ast\zt}\\ 
    & \quad - L\zt\sxerro\lt - M\zt \spdz{\sxerro\lt}\\
    \sbcn{\sxerro} & = -l_0\t\sxerro\nt\\
    \sbce{\sxerro} & = -l_\ell\t\sxerro\lt.
  \end{split}
\end{align}
Similar to the control design subsequently a backstepping approach is
utilized in terms of 
\begin{align}\label{eq:sec4:bstepTrnsfObsv}
  \xerro = \wtsymb\zt - \integral{l\zst \wtsymb\st}{s}{0}{z}
\end{align}
with the kernel $l\zst$ defined on $ \zst \in
\mathcal{T}_{l}(\ell):=\ofB{\zst \in \R^2 \times \Rtn \mid
  s\in[0,\ell], z\in[0,s]}$ to map \eqref{eq:sec4:obsvErrSys} into the
target dynamics
\begin{align}\label{eq:sec4:tarSysObsv}
  \begin{split}
    \sdcr[{&}]{\wtsymb\zt}{a}{0}{0}-\nu\t \wtsymb\zt\\ 
    \spdt{\wtsymb}\nt &= -\nu\t \wtsymb\nt\\
    \spdt{\wtsymb}\lt &= -\nu\t \wtsymb\lt\\
    \wtsymb\ztn &= \wtsymb_0\z \,.
  \end{split}
\end{align}
Proceeding as in Section \ref{sec:bckstStabCtrl} the kernel equations
are obtained as 
\begin{align}\label{eq:sec4:intKernPDEObsv}
  \begin{split}
    \spdt{l\zst} &= a\spdz[2]{l\zst} - a\spd[2]{l\zst}{s}\\
    &\qquad+(\gamma\zt+\nu\t){l\zst}\\
    l\sst &= \frac{1}{2a}\intzntos{\ofB{\gamma\zt+\nu\t}}\\
    l\lst &= 0
  \end{split}
\end{align}
implying the weights 
\begin{align}
  \begin{split}\label{eq:sec4:obsvGainsMap}
    L\zt &= -a(l\zlt l\llt + \spd{l\zlt}{s}) \\
    M\zt &=  a l\zlt.
  \end{split}
\end{align}
Evaluation of \eqref{eq:sec4:bstepTrnsfObsv} at the boundaries
$z\in\{0,\ell\}$ taking into account \eqref{eq:sec4:obsvErrSys},
\eqref{eq:sec4:tarSysObsv}, and \eqref{eq:sec4:intKernPDEObsv} leads to 
\begin{align}
  \label{eq:sec4:obsvGainsMap:bc}
  \l_0\t=\l_\ell\t=\nu\t.
\end{align}
The solution of the PDE \eqref{eq:sec4:intKernPDEObsv} and the strong
solution properties can be determined as in Section
\ref{sec:bckstStabCtrl}. Similarly the stability analysis of Section
\ref{sec:fbCtrlStabAnlys} carries over to verify the exponential
convergence of the observer error dynamics \eqref{eq:sec4:obsvErrSys} with
\eqref{eq:sec4:obsvGainsMap}, \eqref{eq:sec4:obsvGainsMap:bc} to the
zero state. The stability of the combined observer and feedback
control structure follows by making use of the separation principle
in view of the cascaded structure
\citep{frihauf_leader-enabled_2011,meurer_control_2013}.


%
\section{Simulations results}\label{sec:simStuds}
%
Simulation results are presented for the proposed
trajectory planning and tracking control scheme for the formation control
of a multi-agent system. 
%
\subsection{Relocating formation profiles}\label{subsec:relocation}
\begin{figure}[!h]
  \centering
  \includegraphics[width=\linewidth]{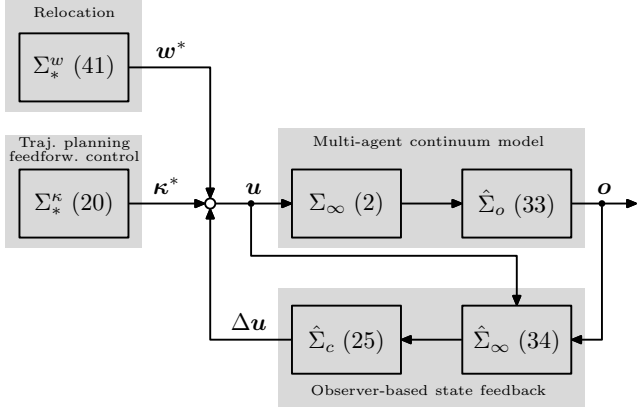}
  \caption{Block diagram of the tracking control scheme with
    profile relocation. Herein the abbreviations
    $\bs{w}^\ast\t=[w_0^{\ast}\t,w_\ell^{\ast}\t]$, 
    $\bs{\kappa}^{\ast}\t=[(\bs{\kappa}_0^{\ast}\t)^{T},(\bs{\kappa}_\ell^{\ast}\t)^T]^T$
    and $\Delta\bs{u}\t=[\Delta u_0\t,\Delta u_\ell]^T$ are used. 
  }
  \label{fig:blockdiagram}
\end{figure}

%
By construction formation profiles \eqref{eq:sec2:stdyStAnsatz1} are
typically arranged around some centre point in the $(x^1,x^2)$-plane,
mostly about the origin. To achieve a relocation of the profile an
exogenous system can be added, e.g., in terms of the heat 
equation 
\begin{align}\label{eq:sec2:exoSysPDE}
  \begin{split}
    \sdcr[{&}]{\xexo}{a}{0}{0}\,,\\
    \sbcn{\sxexo} &=w_{0}\t,\quad \sbcl{\sxexo} = w_{\ell}\t\,,\\
    \sxexo(\zsymb,\tsymb_0) &= x_{\mathrm{e},0}\z.
  \end{split}
\end{align}
In view of the trajectory planning results from Section
\ref{sec:flatnessFF} it can be in a straightforward way deduced that
finite time transition between steady state solutions of
\eqref{eq:sec2:exoSysPDE}  can be realized by interpreting the
boundary values $w_{0}\t=w_0^{\ast}\t$ and
$w_{\ell}\t=w_\ell^{\ast}\t$ as feedforward controls and 
suitably assigning their temporal path, e.g., by exploiting again the
flatness property of \eqref{eq:sec2:exoSysPDE}. Note that these steady
states are given in the form $\bar{x}_\mathrm{e}\z= p_0 + p_1z$ with
$p_0=\bar{x}_\mathrm{e}(0)$ and
$p_1=(\bar{x}_\mathrm{e}(\ell)-\bar{x}_\mathrm{e}(0))/\ell$ with the
value $\bar{x}_\mathrm{e}(0)$, $\bar{x}_\mathrm{e}(\ell)$ being freely
assigned.
\begin{rem}
  Similar to the multi-agent system model
  \eqref{eq:sec2:problem} with
  $c\zt$ enabling the information propagation adding the exogenous system
  \eqref{eq:sec2:exoSysPDE} allows for a decentralised distribution of
  the relocation profile.
  For this, the state of any agent at
  $z\in[0,\ell]$ is described in terms of three states, i.e.,
  $[x,c,x_e]\zt$, or six states, i.e.,
  $[x^{1},c^{1},x_e^{1},x^{2},c^{2},x_e^{2}]\zt$,
  respectively, when taking into account the planar motion in the
  $(x^{1},x^{2})$-domain. 
\end{rem}
With \eqref{eq:sec2:exoSysPDE} manipulated only by means of the
feedforward controls $w_{0}\t$ and $w_{\ell}\t$ providing the
open-loop state evolution $\xexo$ the tracking error fulfils 
\begin{align*}
  \xerrc &= \x-\xdes\\
         &= \x+\xexo - (\xdes+\xexo)\\
         &= \xshft - \xdes[_s].
\end{align*}
As a result, the feedback control and the observer design without any
modification apply in the relocation setting. Hence, subsequently no
distinction is made between $\x$ and $\xshft$. The resulting
control-loop is shown in the block diagram in
Fig. \ref{fig:blockdiagram}. 
%
%
\subsection{Communication topology}\label{subsec:communication}
%
The transfer from the continuum description to the discrete
formulation is obtained by using a finite difference discretization
for the arising PDE models \eqref{eq:sec2:problem},
\eqref{eq:sec4:obsvSys} and \eqref{eq:sec2:exoSysPDE}. For the $x$- and $c$-dynamics
(and similarly for the $x_e$- or $x_s$-dynamics) this results in the
formulation \eqref{eq:sec2:discrete:all} taking into account Proposition
\ref{prop:sec2:discrete_vs_cont} and its proof. This refers to
either time-scaling for fixed value of $\ell$ or vice
versa. Subsequently, the latter is chosen by keeping $t$ unscaled and
setting $\ell=N$ given $N+1$ agents so that $\Delta z=1$. For this
choice the reader is also referred to Remark \ref{rem:sec6:cfl}.

The observer \eqref{eq:sec4:obsvSys} requires at least the
availability of the values $ \x $ at $z\in\{0,\ell\}$ and $ \spdz\x $
at $ \zsymb=\ell $.
Since the observer state is in the considered setting only used
to evaluate the feedback controller $\fbl$ defined in
\eqref{eq:sec4:leaderLaw} it is reasonable to evaluate the discretized
observer equations at the node $z=\ell$. Alternatively, a distributed
evaluation is possible provided that any node has access to the
boundary values. The arising integral in \eqref{eq:sec4:leaderLaw} is
approximated using the Simpson's rule. 
%
\subsection{Simulation studies}
%
\begin{figure*}[tb]
  \begin{center}
    \subfigure[]
    [Observer-based 2DOF control with $\cc$ assigned  explicitly.]{%
      \includegraphics[width=5.5cm]{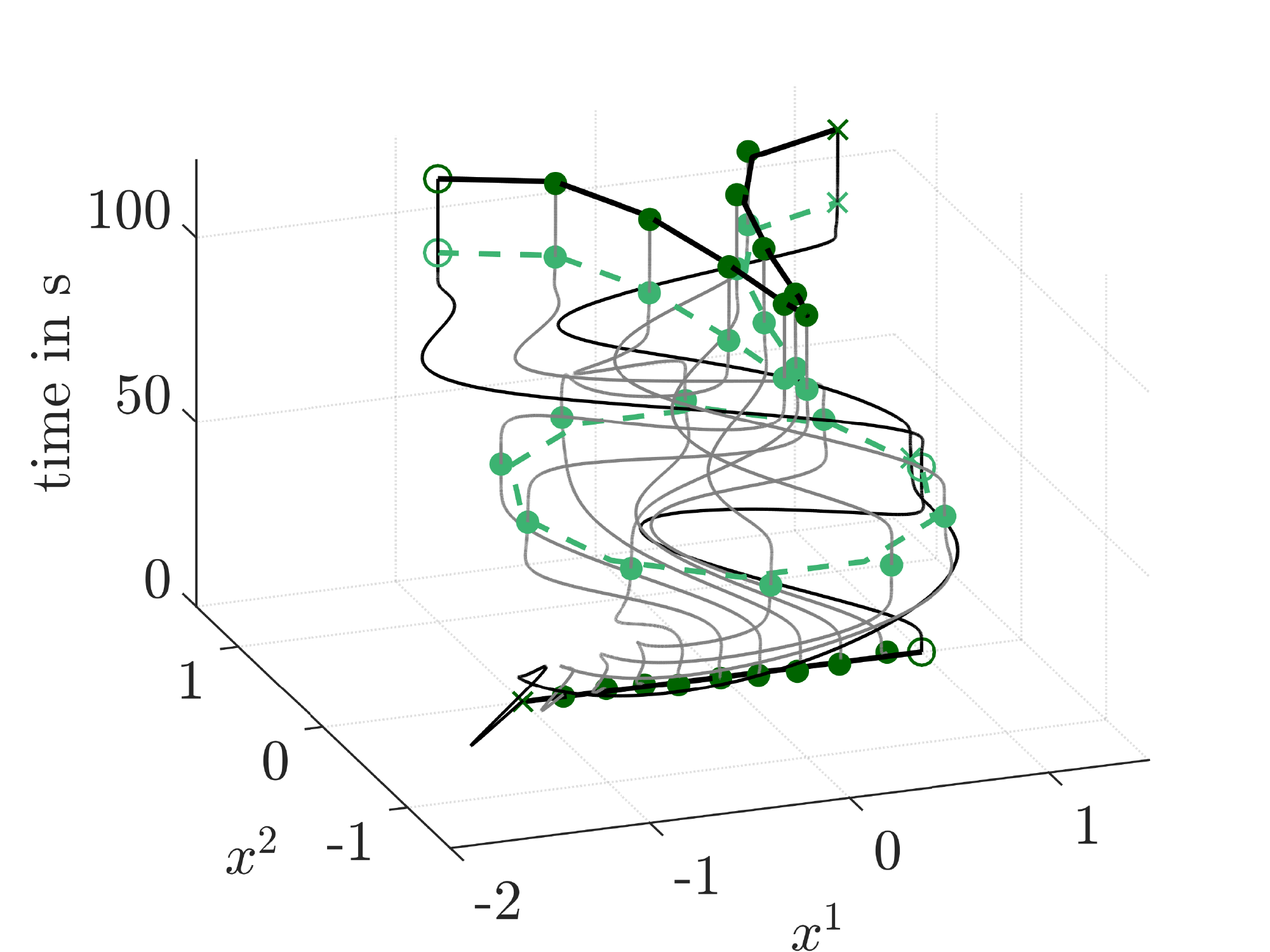}\label{fig:sec5:transObsv}}
    \quad
    \subfigure[]
    [Observer-based 2DOF control with $\cc$ as in (2) but computed in a distributed fashion by PDE \eqref{eq:sec2:problemPDE:c}.]{%
      \includegraphics[width=5.5cm]{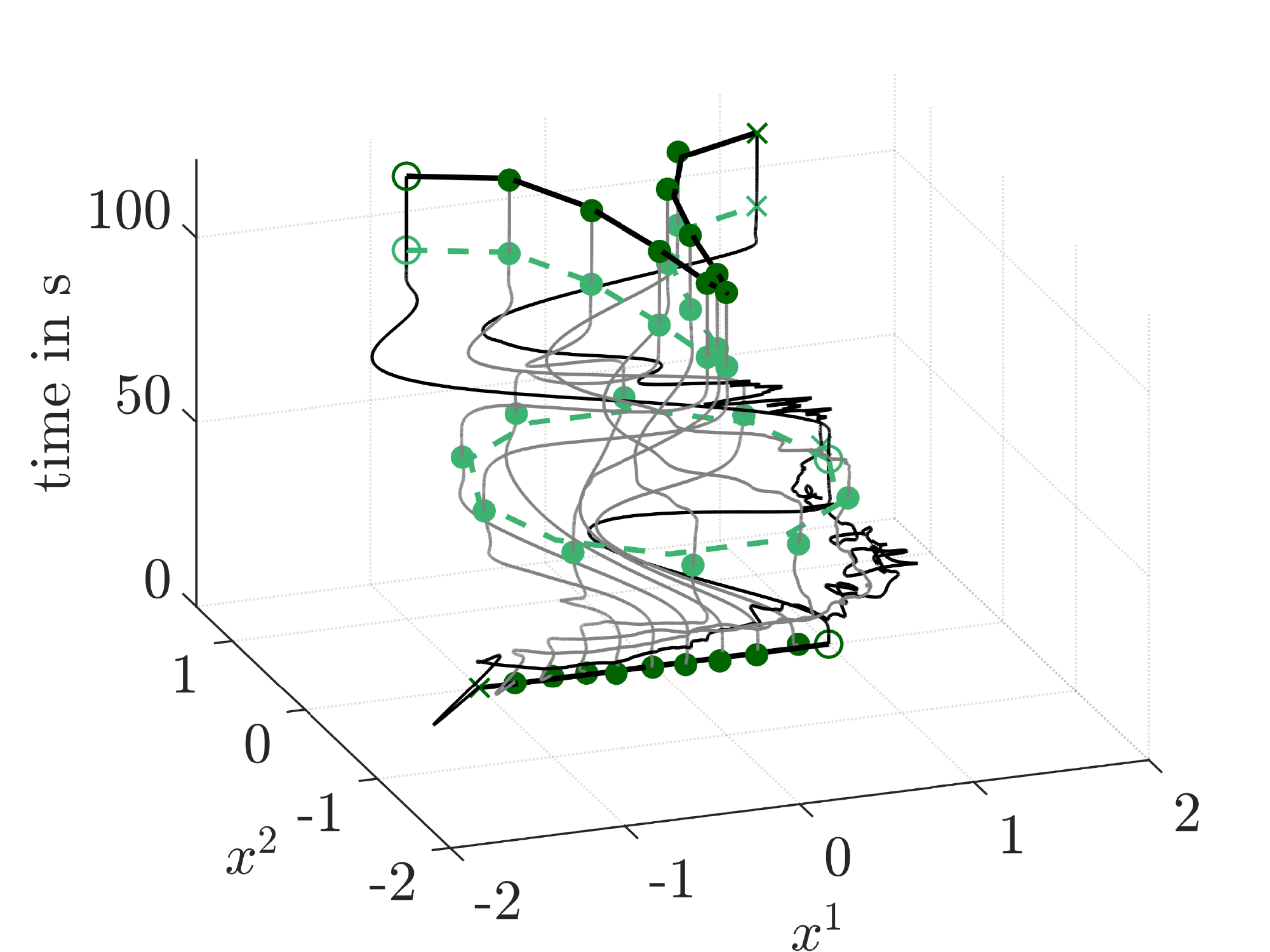}\label{fig:sec5:transObsvCexo}}
    \quad
    \subfigure[]
    [Previous scenario \subref{fig:sec5:transObsvCexo} but with
    relocation of final formation center to virtual centre $(1,0)$.]{%
      \includegraphics[width=5.5cm]{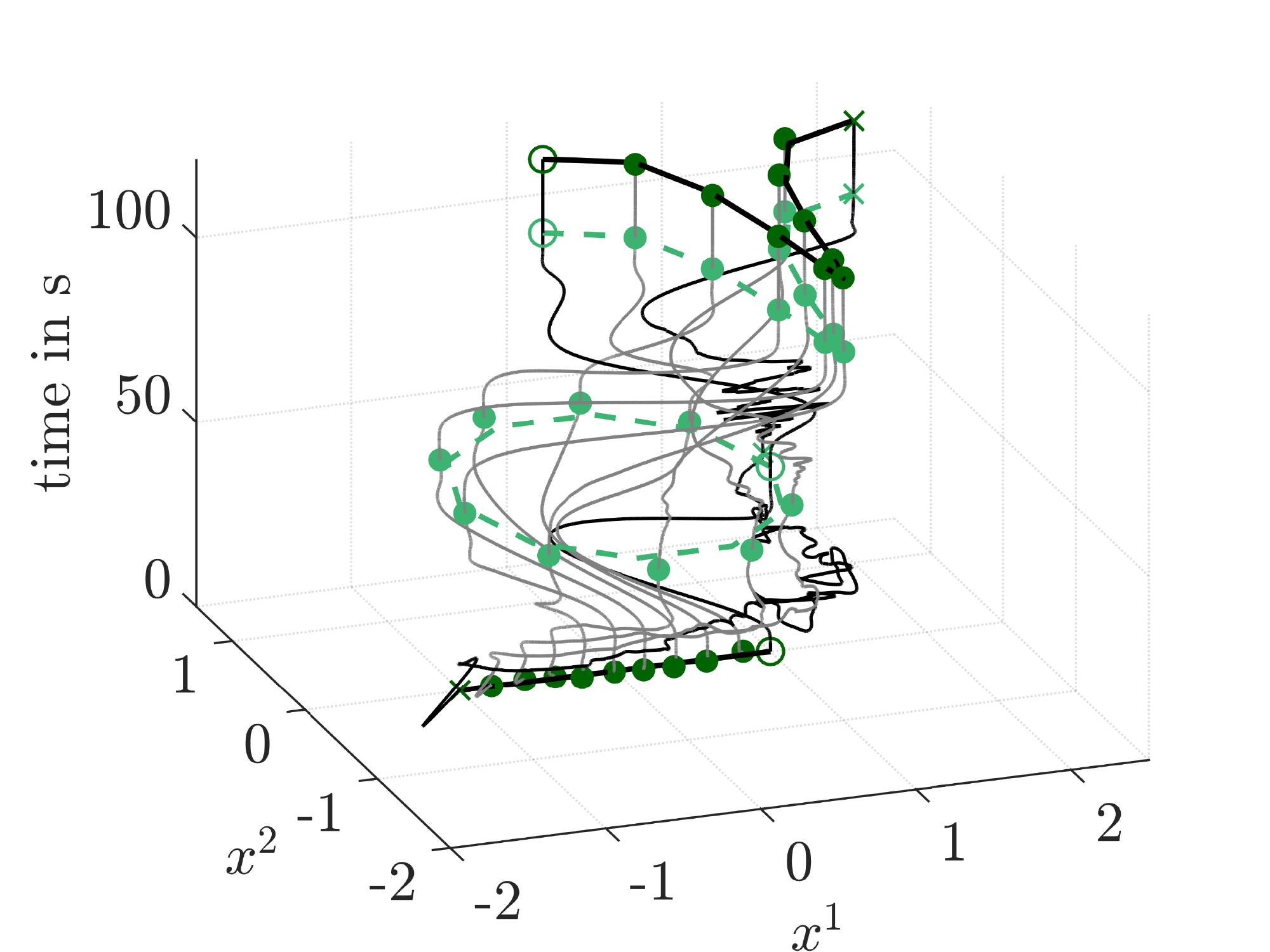}\label{fig:sec5:transObsvCXexoRelo}}
    \\[1ex]
    \begin{minipage}{5.5cm}
      \centering
      \subfigure[][$L^2$-norm of tracking error in \subref{fig:sec5:transObsv}.]{%
        \includegraphics{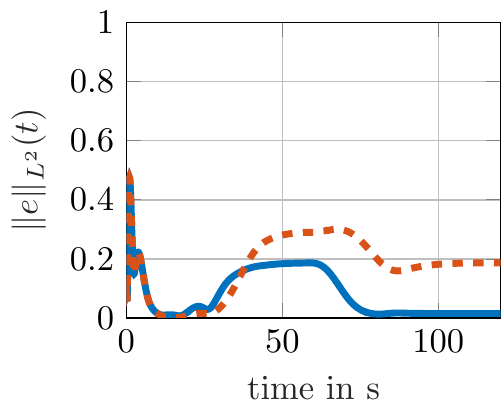}\label{fig:sec5:transObsv_xerrc}}
    \end{minipage}
    \begin{minipage}{5.5cm}
      \centering
      \subfigure[][$L^2$-norm of tracking error in \subref{fig:sec5:transObsvCexo}.]{%
        \includegraphics{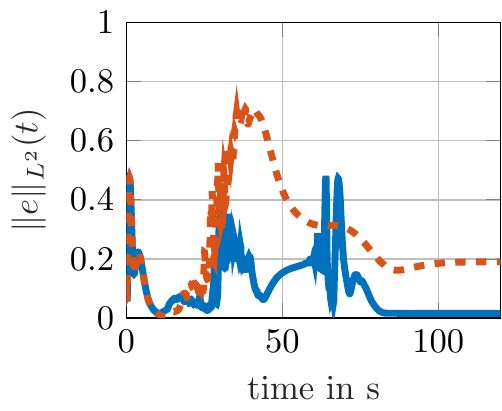}\label{fig:sec5:transObsvCexo_xerrc}}
    \end{minipage}
    \begin{minipage}{5.5cm}
      \centering
      \subfigure[][$L^2$-norm of tracking error in \subref{fig:sec5:transObsvCXexoRelo}.]{%
        \includegraphics{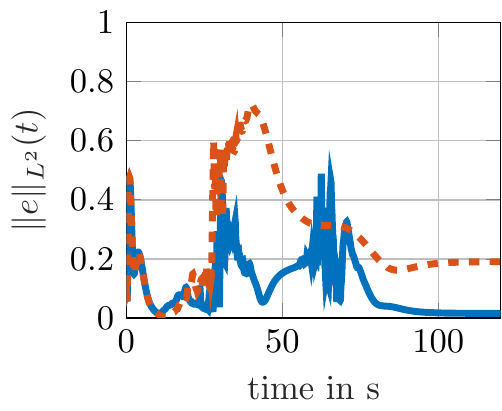}\label{fig:sec5:transObsvCXexo_xerrc}}
    \end{minipage}
  \end{center}
  \caption{Consecutive finite time transitions of $N+1=11$
    agents from a line formation to an intermediate circle and to a
    final gull-like formation. The symbols $\circ$ and $\times$ refer to the
    leader agents at $z=0$ and $z=\ell$ while $\bullet$ denote
    follower agents. Black lines indicate the initial and the reached
    final formation, green dashed lines illustrate desired
    intermediate formation profiles. Top row: spatial-temporal
    transition paths from the overlay of $\sx^1\zt$ and $\sx^2\zt$;
    bottom row: tracking errors with
    $\|\sxerrc^{1}\|_{L^2}\t$ in blue and $\|\sxerrc^{2}\|_{L^2}\t$ in
    red (dashed).}
  \label{fig:sec7:simResults}
\end{figure*}
Two transitions are performed in each of the three simulation
scenarios for $N+1=11$ agents ($\ell=10$) illustrated in
\figref{fig:sec7:simResults}. In all studies
\subref{fig:sec5:transObsv}-\subref{fig:sec5:transObsvCXexoRelo} the
agents start with the same line formation at $ \tsymb=0 $ and then
move to a circular formation. During the second transition the
deployments change from a circle to a gull-like shape.
Note that the line formation is stable by design while the circular
formation is open-loop unstable for both coordinates. For the
gull-like formation the $\sx^2$-coordinate remains open-loop
unstable but the coefficient $c^{1}\zt$ in the PDE governing $ \sx^1\zt $
becomes negative, i.e., it changes from an open-loop unstable to a stable
formation profile in $\sx^1\zt$.
In Fig. \ref{fig:sec5:transObsvCXexoRelo} the formation additionally
moves to the virtual centre $ (1,0) $ using the relocation approach
proposed in Section \ref{subsec:relocation}. Each of the two
transitions lasts $ 50\,\si{\second} $ and the entire simulation time
is set to $120\,\si{\second} $.
The diffusion coefficient in \eqref{eq:sec2:problem} is set to $a^i=1$
for all scenarios. These differ in their problem setup:
\begin{itemize}
\item Fig. \ref{fig:sec5:transObsv} shows the transitions obtained
  with the 2DOF controller with $ \cc[^i]=\sc^{\ast i}\t $ directly
  assigned as a Gevrey function, equivalently to
  \eqref{eq:sec3:FltOutp:j} with steady state data according to
  \tabref{tab:sec7:SimParamters}. The 
  position data for the evaluation of \eqref{eq:sec4:leaderLaw}
  is estimated by the developed state observer.
\item The scenario in Fig. \ref{fig:sec5:transObsvCexo} is identical
  to the previous study but implements the PDE
  \eqref{eq:sec2:problemPDE:c} for the distributed
  computation of the reaction coefficient $\cc$ by imposing $d^{i}\zt=0 $,
  $b^{i}=(N+1)/2=11/2$ and $\vv[_0^i]=\vv[_0^{\ast i}]$,
  $\vv[_\ell^i]=\vv[_\ell^{\ast i}]$ according to the second elements
  of the vectors in \eqref{eq:sec3:fdfdterms}. Note that $b^{i}>a^i$
  is chosen to obtain a faster convergence of $\cc[^{i}]$ compared to $\x[^{i}]$. 
\item In Fig. \ref{fig:sec5:transObsvCXexoRelo} additionally the
  information of the centre point is propagated in a decentralized way
  through the agent topology by means of the relocation procedure
  using \eqref{eq:sec2:exoSysPDE} for $w_{0}\t$, $w_{\ell}\t$ designed
  as feedforward terms following the procedure of Section \ref{sec:flatnessFF}.
\end{itemize}
The corresponding steady state parameters can be studied from
\tabref{tab:sec7:SimParamters}. Controller and observer gains are assigned as
$\mu\t=0.5$ and $ \nu\t=0.6 $, respectively, for both coordinates. 
\begin{table}[tb]
  \caption{Desired steady state transition parameters to be
    taken at time instances $t=\tau_1$ and $t=\tau_2$.}
  \begin{tabular}{lcccccc}
    \toprule
    &\multicolumn{3}{c}{$\sx^1\zt $}&\multicolumn{3}{c}{$\sx^2\zt $}\\[-0.4em]
    $ \tsymb $: & $ 0 $ & $ \tau_1 $ & $ \tau_2 $& $ 0 $ & $ \tau_1 $ & $ \tau_2 $\\
    \cmidrule(l){2-4}\cmidrule(l){5-7}
    $\sxavr_0^{\ast}$: & $ -1 $ & $ 1 $ & $ -1 $& $ 0 $ & $ 0 $ & $ 1 $\\[-0.2em]
    $\sxavr_\ell^{\ast}$: & $  1 $ & $ 1 $ & $ 1 $ & $ 0 $& $ 0 $ & $ 1 $\\[-0.4em]
    $\bar{\sc}^{\ast} $:   & $ 0 $ & $ (2\pi/\ell)^2 $ & $  -(7/\ell)^2 $ & $ 0 $ & $ (2\pi/\ell)^2 $ & $ (2\pi/\ell)^2$\\[-0.4em]
    $\bar{x}_\mathrm{e}^{\ast} $:   & $ 0 $ & $ 0 $ & $ 1 $ & $ 0 $ & $ 0 $ & $ 0 $\\
    \bottomrule
  \end{tabular}\vspace*{-0.25em}
  \label{tab:sec7:SimParamters}
\end{table}
To test the robustness of the approach towards the real-time
application in Section \ref{sec:expResults} the following
deviations from the nominal case are introduced into any simulation
scenario: 
\begin{itemize}
\item First, the sample time of the boundary control inputs and the
  observer is set to $ t_s^o=10~\si{\milli\second} $, while
  the update interval of the exogenous system
  \eqref{eq:sec2:exoSysPDE} and the subsystem for $ \cc $ is specified
  as $ t_s^e=20~\si{\milli\second} $.
\item Second, the propagation of information of the
  exogenous system \eqref{eq:sec2:exoSysPDE} and the $ \cc$-subsystem
  require (wireless) communication messages between the agents. For
  the simulations information drop-outs are induced to model
  the loss of messages. The consequences of these drop-outs are randomly
  lagging values of $ \cc $ and $ \xexo $ for the followers.
\item Third, the multi-agent system does not start in its intended line
  formation but a random initial control and observation
  error is induced. The error is limited to $ \pm5~\si{\percent} $ of the
  formation amplitude, e.g., here $\abs\xerro\leq0.05 $ given the
  circle radius is $ 1 $.
\end{itemize}
Under these circumstances the performance of the control concept can
be evaluated by studying
Figs. \ref{fig:sec5:transObsv_xerrc}-\ref{fig:sec5:transObsvCXexo_xerrc} 
which show the $ L^2$-norm of the 
tracking errors $ \sxerrc^{1}\zt $ and $\sxerrc^{2}\zt$ for the three
simulation scenarios. 
Despite the imposed errors the 2DOF
control concept is in any studied case capable of realizing stable
transitions between the different formations profiles.
The introduction of the $c$-subsystem \eqref{eq:sec2:problemPDE:c},
\eqref{eq:sec2:c:cont:leader} to distribute parameter information
involving simulated information drop-outs and the relocation of the
centre point from $(0,0)$ to $(1,0)$ as expected yield slightly
larger tracking errors during transient behavior.


%
\section{Experimental results}\label{sec:expResults}
%
Experimental results are presented from a
laboratory test rig at the Chair of Control, Kiel University. To the
best knowledge of the authors this represents the first real-time
implementation of the backstepping methodology based on parabolic 
PDEs for the formation control of multi-agent systems using continuum models.
\subsection{Multi robot test rig}\label{sec:testRig}
\begin{figure}[htb]
  \begin{minipage}{0.15\textwidth}
    \includegraphics[width=\textwidth]{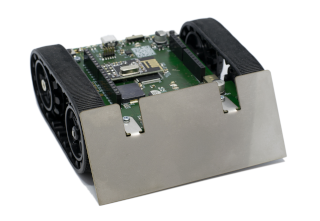}
  \end{minipage}
  \begin{minipage}{0.8\textwidth}
    \tiny\ttfamily{
      \begin{tabular}{ll}
        \toprule
        Type & Features\\
        \midrule
        Processor & ARM Cortex-M4F \@ $ 120 $MHz\\
        RAM       & $ 2\times64 $kB \\
        Flash     & $ 512 $kB\\
        Communication & USB, nRF$ 24 $, Bluetooth\\
        Motors    & $ 2 $ DC motors with $ 75:1 $ gearbox\\
        \multirow{3}{*}{Periphery} & $ 6  $ axis IMU, $ 2 $ LEDs, Buzzer, \\
             & Magentic quadrature encoders,\\
             & IR sensors, Arduino header, etc.\\
        Dimensions & approx. $ 10 $cm $ \times $ $ 10 $cm $ \times $ $ 4 $cm\\
        \bottomrule
      \end{tabular}}
  \end{minipage}
  \normalfont
  \caption{The agent: a wheeled caterpillar robot \citep{styger_2016}.}
  \label{fig:sec6:sumoRobot}
\end{figure}

\begin{figure}[htb]
  \centering
  \includegraphics[width=0.4\textwidth]{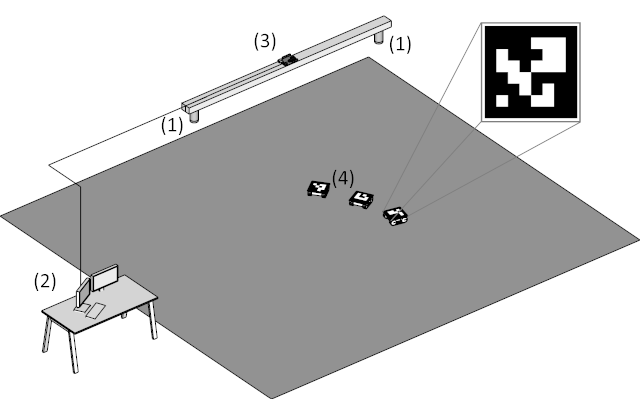}\\
  \caption{Basic scheme of the mobile robot test rig.}
  \label{fig:sec6:testRigScheme}
\end{figure}

Basically the multi-agent system is built upon small caterpillar
robots which are shown \figref{fig:sec6:sumoRobot}, where in addition
the basic features of the used robot are listed to give an impression
of the  available computational power and memory capacity.
The real-time implementation of the 2DOF control concept introduced in
Sections \ref{sec:flatnessFF} and \ref{sec:trckngCtrl} demands to access the
position of each robot agent in the two dimensional plane either by
measurement or by using the state observer.
To address this a suitable hard- and software environment has been set
up to perform controlled transitions between different formations
including their relocation. The used environment is schematically illustrated in
\figref{fig:sec6:testRigScheme} and basically consists of the four
main subsystems:
\begin{enumerate}
\item[(1)] Ceiling-mounted camera or a camera system for optical position
  detection.
\item[(2)] Computer for OpenCV
  application \citep{opencv}.
\item[(3)] Development board with a radio module.
\item[(4)] Caterpillar robots equipped with AruCo codes \citep{garrido-jurado_automatic_2014}.
\end{enumerate}
In the test environment a ceiling-mounted camera is used for global
position measurement and subsequently emulates the induced
communication topology of the 
multi-agent system. For this, a work station runs an image processing
application which uses OpenCV and includes the so-called AruCo
library. The latter is used to detect the individual AruCo codes,
which are fixed on top of each agent
\citep{garrido-jurado_automatic_2014}. The image data is processed and
is sent via a serial interface to an electronic development board,
which is equipped with a radio module. The electronic board runs a software
which broadcasts messages with the position information of all agents
via radio to the caterpillar robots. The caterpillar robots, serving
as agents, are equipped with two DC motors
and a radio module. From the broadcast each robot only extracts its specific
position information according to the underlying communication
topology, which is induced by the input protocol
\eqref{eq:sec2:x:discrete:ode}.
\begin{rem}
  It should be emphasised that the used caterpillar robot represents a
  non-holonomic system. Its kinematic model has the form
  $\dot{\sx}^1_r\t = \vsymb_r\t\cos(\phi_r\t)$, $\dot{\sx}^2_r\t =
  \vsymb_r\t\sin(\phi_r\t)$, $\dot{\phi}_r\t = \omega_r\t$
  with $ v_r\t $ the translational velocity and $ \omega_r\t $ the
  angular velocity defining the robot orientation $ \phi_r\t $ in the
  2D plane. Obviously the model has to satisfy the non-holonomic
  constraint $\dot{\sx}^2_r\t \cos(\phi_r\t)= \dot{\sx}^1_r\t\sin(\phi_r\t)$.
  This behaviour, induced by the robot kinematics, somewhat
  counteracts the modeling assumptions, where the agents are in
  principle represented as ideal mass points. Moreover, it provides a
  significant challenge for the developed 2DOF controllers to compensate this
  difference hence imposing a benchmark for robustness
    analysis. 
\end{rem}
%
\subsection{Test scenario}
\begin{figure*}[!t]
  \centering
  \includegraphics[width=5.5cm]{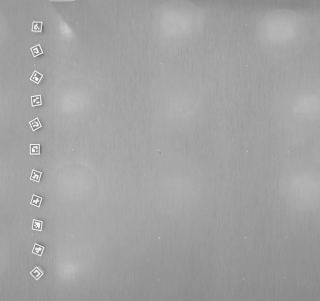}
  \includegraphics[width=5.5cm]{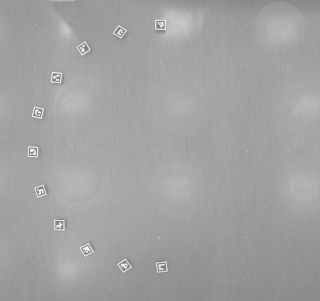}
  \includegraphics[width=5.5cm]{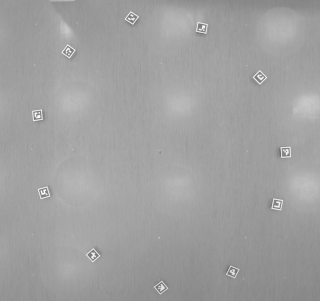} 
  \caption{Snapshots of the transition using the ceiling-mounted camera system.}
  \label{fig:photos:transitions}
\end{figure*}

\begin{figure*}[tb]
 \begin{center}
    \begin{minipage}{0.49\linewidth}
      \subfigure[][Flatness-based feedforward control with backstepping-\\based error feedback.]{%
        \includegraphics{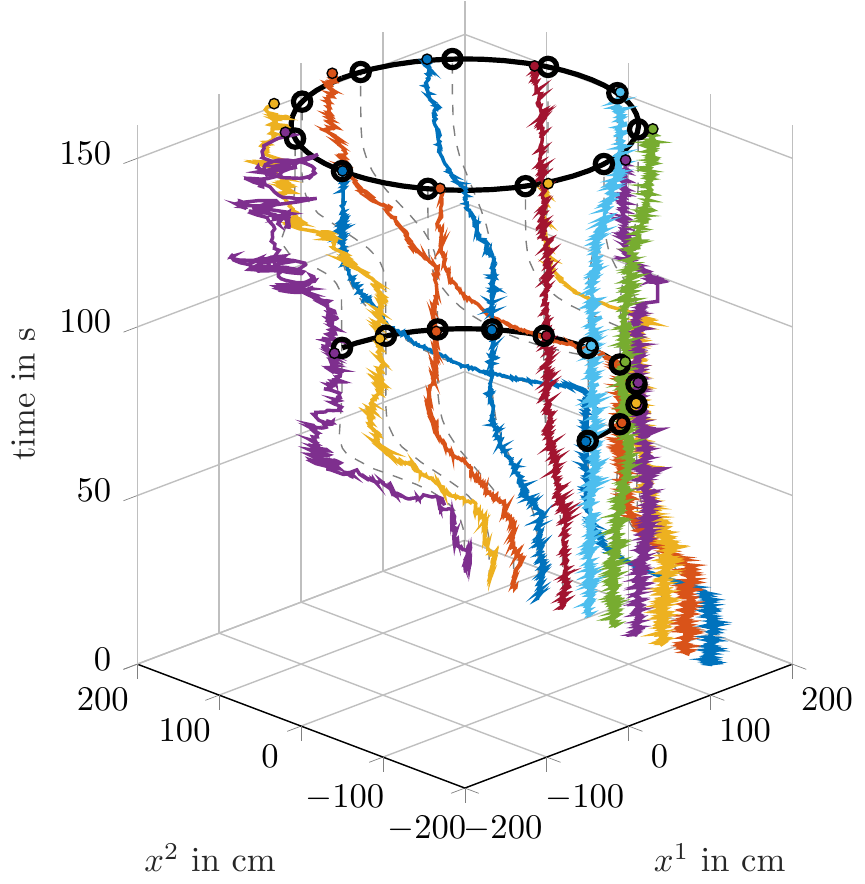}\label{fig:sec6:circle_bckstp_3D}}
    \end{minipage}
    \begin{minipage}{0.49\linewidth}
      \subfigure[][Flatness-based feedforward control with proportional error feedback.]{%
        \includegraphics{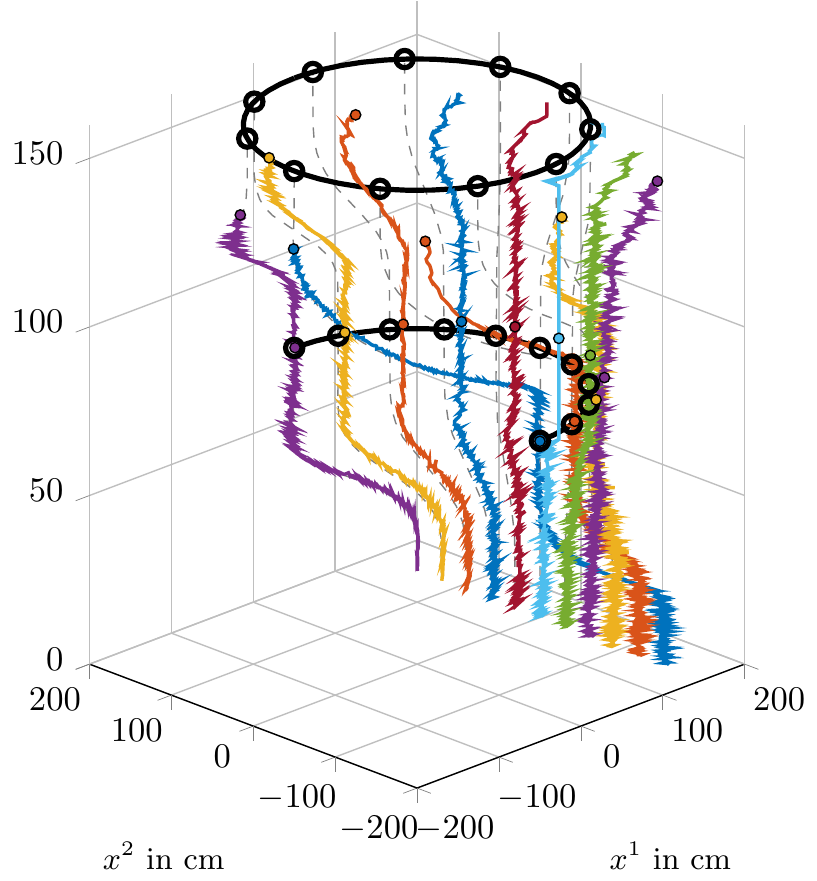}\label{fig:sec6:circle_PT1_3D}}
    \end{minipage}\\[1ex]
    \begin{minipage}{0.49\linewidth}
      \subfigure[][Mean distance error for \subref{fig:sec6:circle_bckstp_3D}.]{%
        \includegraphics{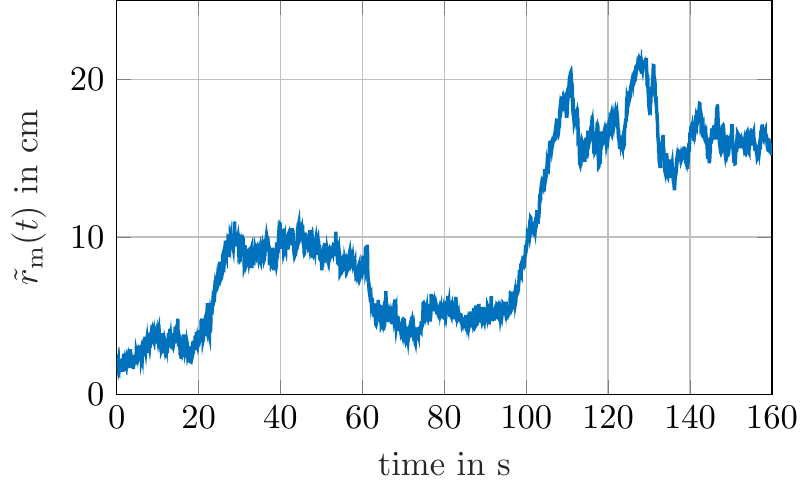}\label{fig:sec6:circle_bckstp_meanErr_2D}}
    \end{minipage}
    \begin{minipage}{0.49\linewidth}
      \subfigure[][Mean distance error for \subref{fig:sec6:circle_PT1_3D}.]{%
        \includegraphics{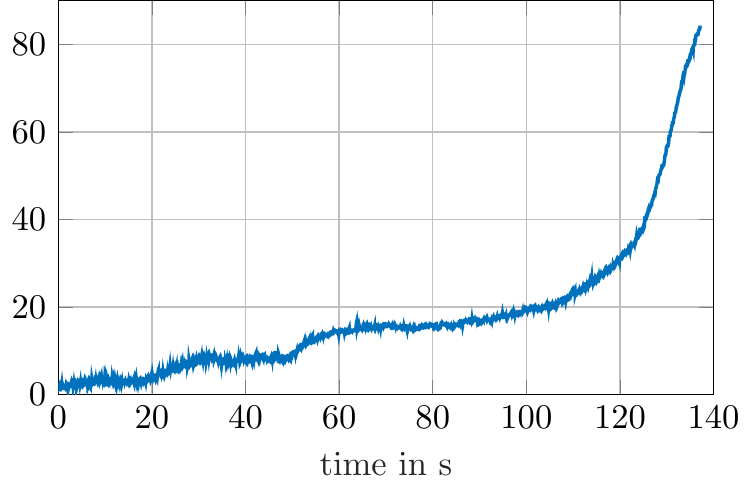}\label{fig:sec6:circle_pt1_meanErr_2D}}
    \end{minipage}

  \caption{Experimental results for the test
    scenario. \subref{fig:sec6:circle_bckstp_3D} shows the
    spatial-temporal evolution of the transition with 2DOF controller
    including a backstepping-based feedback term; for
    \subref{fig:sec6:circle_PT1_3D} the 2DOF controller is evaluated
    with a proportional error feedback;
    \subref{fig:sec6:circle_bckstp_meanErr_2D} and
    \subref{fig:sec6:circle_pt1_meanErr_2D} show the mean distance
    error computed according to \eqref{eq:sec7:errMeanDist}.}
  \label{fig:sec6:testResults}
\end{center}
\end{figure*}

%
The experimental results for $N+1=11$ robots are based on the follower protocol
\eqref{eq:sec2:x:discrete:ode} and the leader protocol
\eqref{eq:sec2:x:discrete:leader} imposed by the discretization
described in Section \ref{subsec:communication}. The time-variant
reaction term $c_j^{i}\t $, $i=1,2$, $j=0,1,\ldots,10$ is for ease of
implementation configured off-line for each agent. The
synchronisation of the temporal 
evolution of the parameter for all agents is reached through a trigger
signal, which is broadcasted via radio. The implementation of the leader
protocol involves the 2DOF controller 
consisting of the flatness-based \fdfd~term 
\eqref{eq:sec3:fdfdterms} and a measurement-based backstepping
controller according to \eqref{eq:sec4:anchorLaw}, \eqref{eq:sec4:leaderLaw}, respectively.
\begin{rem}\label{rem:sec6:cfl}
  In the numerical simulations it is possible to fix $\ell$ and
  to adjust $N$ so that the discretization stepsize $\Delta z =
  \ell/N$ in principle becomes arbitrarily small for $N\gg
  1$. For a numerically stable integration of the resulting ODEs
  this necessitates to choose a sufficiently small time step $\Delta t$
  for numerical stability. This is no longer possible
  at the experimental setup since the time step is imposed by the
  minimal sampling time, which depends on the used sensor,
  actuation, communication, and processing devices involved in the
  control loop. To address this, $\ell=N$ is chosen both in the
  simulation and the experimental results. Note that this choice is
  in line with exposition in Proposition
  \ref{prop:sec2:discrete_vs_cont}. 
\end{rem}
The obtained experimental results are shown in
\figref{fig:sec6:circle_bckstp_3D} for the twofold transition: first from
the initial line configuration passing through the point $(-150,0)\,\si{cm}$ 
to an intermediate half circle formation of radius $R=150\,\si{cm}$
within the time interval $t\in[0,\tau]$ and secondly to a desired 
circle formation of radius $R$ within the time interval
$t\in(\tau,2\tau]$ (see also Fig. \ref{fig:photos:transitions}). 
Herein, the feedforward term and the integral kernel
\eqref{eq:sec4:intKernPDECtrl} are computed off-line and are
implemented using linear interpolation. The 
parameter setting for the experiment is listed in
\tabref{tab:sec7:testScenParameters}. In view of the 
parameter values and the ansatz \eqref{eq:sec2:stdyStAnsatzCS} the
desired steady state solutions for $ t=2\tau $ can be written
as
\begin{align*}
  \xavr[^{1,\ast}] = -R \cos(\omega \zsymb),~
  \xavr[^{2,\ast}] = -R \sin(\omega\zsymb)
\end{align*}
with $\omega={2\pi}/{(\ell+1)}$. Differing from the simulation
studies before the spatial period $\omega$ of the $\sin$- and
$\cos$-functions is reduced from $2\pi/\ell $ to $2\pi/(\ell+1)$ and
the values for $ \sxavr_0^{i,\ast}$, $\sxavr_\ell^{i,\ast}$, $i=1,2$
are shifted appropriately. Since no explicit collision avoidance
algorithm is used during the transitions this choice of $\omega$
implies that the leader nodes $\zsymb=\{0,\ell\}$ are separated (for
$\omega=2\pi/\ell$ in both leaders will be located at the same point
in the circular formation).
For comparison reasons and to illustrate the performance of the 2DOF
control concept in \figref{fig:sec6:circle_PT1_3D} experimental
results are provided for the combination of the flatness-based
feedforward control with proportional error control at the leader
agents in the form \eqref{eq:sec4:anchorLaw}, i.e., $\fbn= -\mu\t
\sxerrc\nt$ and $\fbl=-\mu\t \sxerrc\lt$ with $\mu\t=0.15$. 
The mean distance error
\begin{align}\label{eq:sec7:errMeanDist}
  \tilde{r}_\mathrm{m}\t &= \frac{1}{N+1}\sum_{j=0}^N \tilde{r}_i\t
\end{align}
with $(\tilde{r}_i\t)^2={(\sx^1_j\t-\sx^{\ast,1}_j\t)^2 +
  (\sx^2_j\t-\sx^{\ast,2}_j\t)^2}.$
between desired and measured position values is shown in
\figref{fig:sec6:circle_bckstp_meanErr_2D} and
\figref{fig:sec6:circle_pt1_meanErr_2D}. 
Analysing the results of \figref{fig:sec6:testResults} clearly reveals
that the 2DOF controller including the backstepping-based error feedback
is able to stabilise the transitions while the simple proportional
error feedback fails and the desired formation falls apart.
\begin{table}[tb]
  \caption{Parameters for the test scenario with $ a=0.5 $, $
    \ell=10 $ and $ \tau=80$ s. Geometric values are given in
    cm.}
  \vspace*{0.5em}
  \begin{tabular}{l p{0.01cm} ccc }
    \toprule
    & $ \tsymb $      & $ 0 $ & $ \tau $ & $ 2\tau $\\
    \cmidrule(l){3-5}
    \multirow{3}{*}{Coord. $ 1 $:}
    &$\bar{x}_0^\ast$    & $ 150$ & $ 0 $ & $ -150 $ \\[-0.2em]
    &$\bar{x}_\ell^\ast$ & $ 150$ & $ 0 $ & $ -150\cos(2\pi\ell/(\ell+1)) $\\[-0.4em]
    &$\bar{c}^\ast$      & $ 0 $ & $ a(\pi/\ell)^2 $ & $ a(2\pi/(\ell+1))^2 $ \\[-0.2em]
    \cmidrule(l){3-5}
    \multirow{3}{*}{Coord. $ 2 $:}
    &$\bar{x}_0^\ast$    & $ -150 $ & $ -150 $ & $ 0 $ \\[-0.2em]
    &$\bar{x}_\ell^\ast$ & $ \phantom{-}150 $ & $ \phantom{-}150 $ & $ -150\sin(2\pi\ell/(\ell+1)) $\\[-0.4em]
    &$\bar{c}^\ast$      &  $ 0 $ & $ a(\pi/\ell)^2 $ & $ a(2\pi/(\ell+1))^2 $\\[-0.2em]
    \bottomrule
  \end{tabular}
  \label{tab:sec7:testScenParameters}
  \normalsize
\end{table}

\newlength\figWidth
\newlength\figHeight
\setlength{\figHeight}{4cm}
\setlength{\figWidth}{7cm}


\section{Conclusion}\label{sec:concl}
Based on a continuum model in terms of coupled PDEs a 2DOF control
concept is developed for the deployment of multi-agent systems into
desired formation profiles. Diffusion-reaction equations are set up to
govern the spatial-temporal agent dynamics in the plane and
simultaneously  enabling us to also distribute (decentralized)
parameter information. 
Based on the PDE model flatness-based trajectory planning is addressed
and combined with backstepping-based state feedback control to achieve
the stable tracking of desired spatial-temporal profiles. For the required
state estimation a backstepping-based Luenberger observer is designed
and integrated into the closed-loop control. With this, finite time
transitions between desired formation profiles, which are determined
as possibly unstable steady state solutions of the governing PDEs, can
be realized. Due to inclusion of time variant parameters this includes
the connection between different families of steady states.
The distribution and propagation of parameter values
between the agents is directly incorporated into the setting in terms
of a feedforward control approach. Furthermore the incorporation of an
exogenous system enables us to achieve the spatial relocation of the
formation. The transfer of the determined controller and estimation
algorithms to the finite-dimensional discrete multi-agent network is
achieved using finite difference discretization. Depending on the PDE
model this may even result in a decentralised implementation and
directly imposes the necessary chain-like communication topology.
Simulations studies show the tracking performance and the robustness
of the concept even in view of rather challenging deviations from the
expected behaviour. These findings are confirmed also in first
experimental results conducted with a small swarm of caterpillar
robots. By means of the developed 2DOF control concept transitions
between different and also unstable (with respect to the considered
PDEs) formation profiles are achieved. To the best knowledge of the
authors the presented results are the first real-time implementation
of the backstepping methodology for parabolic PDEs and the use of
controllers based on continuum models for multi-agent systems.



\begin{ack}                               
The financial support by the Deutsche Forschungsgesellschaft (DFG) in
the individual grant ref. 266006167 is gratefully acknowledged.
The authors would like to thank Prof. Erich Styger from Lucerne 
University of Applied Sciences and Arts for his support concerning
the embedded computing facilities of the used caterpillar robots,
Simon Helling for his help during the implementation of the algorithms
at the experimental set-up, and Dr. Petro Feketa for thoughtful
discussions concerning the well-posedness analysis.
\end{ack}

\bibliographystyle{plainnat}        
\bibliography{main}           

\end{document}